\documentclass[twocolumn]{aastex631}

\usepackage{color}
\usepackage{url}
\shorttitle{X-ray polarization in PG~1553+113}
\shortauthors{Middei et al.}

\newcommand\ixpe{\rm{IXPE}}

\begin{document}

\title{Tracking down the broadband polarimetric properties of PG\,1553+113}



%
%
%
%
%
%


\author[0000-0001-9815-9092]{Riccardo Middei}
\correspondingauthor{Riccardo Middei}
\email{riccardo.middei@inaf.it}
\affiliation{INAF Osservatorio Astronomico di Roma, Via Frascati 33, 00078 Monte Porzio Catone (RM), Italy}

\author[0000-0001-9522-5453]{Svetlana G. Jorstad}
\affiliation{Institute for Astrophysical Research, Boston University, 725 Commonwealth Avenue, Boston, MA 02215, USA}
\affiliation{Saint Petersburg State University, 7/9 Universitetskaya nab., St. Petersburg, 199034 Russia}

\author[0000-0001-7396-3332]{Alan P. Marscher}
\affiliation{Institute for Astrophysical Research, Boston University, 725 Commonwealth Avenue, Boston, MA 02215, USA}

\author[0000-0001-9200-4006]{Ioannis Liodakis}
\affiliation{Institute of Astrophysics, Foundation for Research and Technology-Hellas, GR-71110 Heraklion, Greece}

\author[0000-0003-3613-4409]{Matteo Perri}
\affiliation{INAF Osservatorio Astronomico di Roma, Via Frascati 33, 00078 Monte Porzio Catone (RM), Italy}
\affiliation{Space Science Data Center, Agenzia Spaziale Italiana, Via del Politecnico snc, 00133 Roma, Italy}

\author[0000-0003-3760-1910 ]{Alessandro Maselli}
\affiliation{INAF Osservatorio Astronomico di Roma, Via Frascati 33, 00078 Monte Porzio Catone (RM), Italy}
\affiliation{Space Science Data Center, Agenzia Spaziale Italiana, Via del Politecnico snc, 00133 Roma, Italy}

\author[0000-0002-0712-2479]{Stefano Ciprini}
\affiliation{Space Science Data Center, Agenzia Spaziale Italiana, Via del Politecnico snc, 00133 Roma, Italy}
\affiliation{INAF Osservatorio Astronomico di Roma, Via Frascati 33, 00078 Monte Porzio Catone (RM), Italy}

\author[0000-0001-9226-8992]{Emanuele Nardini}
\affiliation{INAF -- Osservatorio Astrofisico di Arcetri, Largo Enrico Fermi 5, I-50125 Firenze, Italy}

\author[0000-0002-2734-7835]{Simonetta Puccetti}
\affiliation{Space Science Data Center, Agenzia Spaziale Italiana, Via del Politecnico snc, 00133 Roma, Italy}

\author[0000-0002-5614-5028]{Laura Di Gesu}
\affiliation{ASI - Agenzia Spaziale Italiana, Via del Politecnico snc, 00133 Roma, Italy}

\author[0000-0002-3777-6182]{Iv\'an Agudo}
\affiliation{Instituto de Astrof\'{i}sica de Andaluc\'{i}a, IAA-CSIC, Glorieta de la Astronom\'{i}a s/n, 18008 Granada, Spain}

\author[0000-0001-5717-3736]{Dawoon E. Kim}
\affiliation{INAF Istituto di Astrofisica e Planetologia Spaziali, Via del Fosso del Cavaliere 100, 00133 Roma, Italy}

\author[0000-0003-3025-9497]{Ioannis Myserlis}
\affiliation{Institut de Radioastronomie Millim\'{e}trique, Avenida Divina Pastora, 7, Local 20, E–18012 Granada, Spain}
\affiliation{Max-Planck-Institut f\"{u}r Radioastronomie, Auf dem H\"{u}gel 69,
D-53121 Bonn, Germany}

\author[0000-0001-8112-3464]{A. Trindade Falc\~ao}
\affiliation{NASA Goddard Space Flight Center, Code 662, Greenbelt, MD 20771, USA}

\author[0000-0002-3626-5831]{D. \L. Kr\'ol}
\affiliation{Center for Astrophysics $|$ Harvard \& Smithsonian, 60 Garden Street, Cambridge MA 02138, USA}
\affiliation{Astronomical Observatory of the Jagiellonian University, Orla 171, 30-244 Krak\'ow, Poland}

\author[0000-0002-5037-9034]{Lucio A. Antonelli}
\affiliation{INAF Osservatorio Astronomico di Roma, Via Frascati 33, 00078 Monte Porzio Catone (RM), Italy}
\affiliation{Space Science Data Center, Agenzia Spaziale Italiana, Via del Politecnico snc, 00133 Roma, Italy}

\author{Tommaso Aniello}
\affiliation{INAF Osservatorio Astronomico di Roma, Via Frascati 33, 00078 Monte Porzio Catone (RM), Italy}



\author[0000-0002-9328-2750]{Pouya M. Kouch}
\affiliation{Finnish Centre for Astronomy with ESO (FINCA), 20014 University of Turku, Finland}
\affiliation{Department of Physics and  Astronomy, Quantum, Vesilinnantie 5, FI-20014 University of Turku, Finland}

\author[0000-0002-1445-8683]{Kari Nilsson}
\affiliation{Finnish Centre for Astronomy with ESO (FINCA), 20014 University of Turku, Finland}

\author[0000-0002-9155-6199]{Elina Lindfors}
\affiliation{Department of Physics and  Astronomy, Quantum, Vesilinnantie 5, FI-20014 University of Turku, Finland}

\author{Tapio Pursimo}
\affiliation{Nordic Optical Telescope, Apartado 474, E-38700 Santa Cruz de La Palma, Santa Cruz de Tenerife, Spain}
\affiliation{Department of Physics and Astronomy, Aarhus University, Munkegade 120, DK-8000 Aarhus C, Denmark}

\author{Francisco Jos\'e Aceituno}
\affiliation{Instituto de Astrof\'{i}sica de Andaluc\'{i}a, IAA-CSIC, Glorieta de la Astronom\'{i}a s/n, 18008 Granada, Spain}

\author[0000-0003-2036-8999]{V\'{i}ctor Casanova}
\affiliation{Instituto de Astrof\'{i}sica de Andaluc\'{i}a, IAA-CSIC, Glorieta de la Astronom\'{i}a s/n, 18008 Granada, Spain}

\author[0000-0001-6155-4742]{Gabriel Emery}
\affiliation{Instituto de Astrof\'{i}sica de Andaluc\'{i}a, IAA-CSIC, Glorieta de la Astronom\'{i}a s/n, 18008 Granada, Spain}

\author[0000-0002-4131-655X]{Juan Escudero Pedrosa}
\affiliation{Instituto de Astrof\'{i}sica de Andaluc\'{i}a, IAA-CSIC, Glorieta de la Astronom\'{i}a s/n, E-18008 Granada, Spain}
\affiliation{Center for Astrophysics $|$ Harvard \& Smithsonian, 60 Garden Street, Cambridge MA 02138, USA}

\author[0000-0002-4241-5875]{Jorge Otero-Santos}
\affiliation{Instituto de Astrof\'{i}sica de Andaluc\'{i}a, IAA-CSIC, Glorieta de la Astronom\'{i}a s/n, E-18008 Granada, Spain}
\affiliation{Istituto Nazionale di Fisica Nucleare, Sezione di Padova, 35131 Padova, Italy}

\author{Alfredo Sota}
\affiliation{Instituto de Astrof\'{i}sica de Andaluc\'{i}a, IAA-CSIC, Glorieta de la Astronom\'{i}a s/n, 18008 Granada, Spain}

\author{Vilppu Piirola}
\affiliation{Department of Physics and Astronomy, 20014 University of Turku, Finland}

\author[0000-0002-7262-6710]{George A. Borman}
\affiliation{Crimean Astrophysical Observatory RAS, P/O Nauchny, 298409, Crimea}

\author[0000-0002-3953-6676]{Tatiana S. Grishina}
\affiliation{Saint Petersburg State University, 7/9 Universitetskaya nab., St. Petersburg, 199034 Russia}

\author[0000-0002-6431-8590]{Vladimir A. Hagen-Thorn}
\affiliation{Saint Petersburg State University, 7/9 Universitetskaya nab., St. Petersburg, 199034 Russia}

\author[0000-0001-9518-337X]{Evgenia N. Kopatskaya}
\affiliation{Saint Petersburg State University, 7/9 Universitetskaya nab., St. Petersburg, 199034 Russia}

\author[0000-0002-2471-6500]{Elena G. Larionova}
\affiliation{Saint Petersburg State University, 7/9 Universitetskaya nab., St. Petersburg, 199034 Russia}

\author[0000-0002-9407-7804]{Daria A. Morozova}
\affiliation{Saint Petersburg State University, 7/9 Universitetskaya nab., St. Petersburg, 199034 Russia}

\author[0000-0003-4147-3851]{Sergey S. Savchenko}
\affiliation{Saint Petersburg State University, 7/9 Universitetskaya nab., St. Petersburg, 199034 Russia}

\author[0009-0002-2440-2947]{Ekaterina V. Shishkina}
\affiliation{Saint Petersburg State University, 7/9 Universitetskaya nab., St. Petersburg, 199034 Russia}

\author[0000-0002-4218-0148]{Ivan S. Troitskiy}
\affiliation{Saint Petersburg State University, 7/9 Universitetskaya nab., St. Petersburg, 199034 Russia}

\author[0000-0002-9907-9876]{Yulia V. Troitskaya}
\affiliation{Saint Petersburg State University, 7/9 Universitetskaya nab., St. Petersburg, 199034 Russia}

\author[0000-0002-8293-0214]{Andrey A. Vasilyev}
\affiliation{Saint Petersburg State University, 7/9 Universitetskaya nab., St. Petersburg, 199034 Russia}

\author{Alexey V. Zhovtan}
\affiliation{Crimean Astrophysical Observatory RAS, P/O Nauchny, 298409, Crimea}

\author{Petra Benke}
\affiliation{GFZ Helmholtz Centre for Geosciences, Telegrafenberg, 14476, Potsdam, Germany}
\affiliation{Max-Planck-Institut f\"{u}r Radioastronomie, Auf dem H\"{u}gel 69, D-53121 Bonn, Germany}

\author[0009-0003-8342-4561]{Lena Debbrecht}
\affiliation{Max-Planck-Institut f\"{u}r Radioastronomie, Auf dem H\"{u}gel 69, D-53121 Bonn, Germany}

\author{Julia Eich}
\affiliation{Julius-Maximilians-Universit\"{a}t W\"{u}rzburg, Institut f\"{u}r Theoretische Physik und Astrophysik, Lehrstuhl f\"{u}r Astronomie, Emil-Fischer-Stra{\ss}e 31, 97074 W\"{u}rzburg, Germany}

\author[0000-0001-7112-9942]{Florian Eppel}
\affiliation{Max-Planck-Institut f\"{u}r Radioastronomie, Auf dem H\"{u}gel 69, D-53121 Bonn, Germany}
\affiliation{Julius-Maximilians-Universit\"{a}t W\"{u}rzburg, Institut f\"{u}r Theoretische Physik und Astrophysik, Lehrstuhl f\"{u}r Astronomie, Emil-Fischer-Stra{\ss}e 31, 97074 W\"{u}rzburg, Germany}

\author[0000-0002-5726-5216]{Andrea Gokus}
\affiliation{Physics Department and McDonnell Center for the Space Sciences, Washington University in St. Louis, MO, 63130, USA}

\author{Steven H\"{a}mmerich}
\affiliation{Dr. Karl-Remeis Sternwarte and Erlangen Centre for Astroparticle Physics, Friedrich-Alexander Universit\"at Erlangen-N\"urnberg, Sternwartstr.~7, 96049 Bamberg, Germany}

\author[0009-0009-7841-1065]{Jonas He\ss d\"orfer}
\affiliation{Max-Planck-Institut f\"{u}r Radioastronomie, Auf dem H\"{u}gel 69, D-53121 Bonn, Germany}
\affiliation{Julius-Maximilians-Universit\"{a}t W\"{u}rzburg, Institut f\"{u}r Theoretische Physik und Astrophysik, Lehrstuhl f\"{u}r Astronomie, Emil-Fischer-Stra{\ss}e 31, 97074 W\"{u}rzburg, Germany}

\author[0000-0001-5606-6154]{Matthias Kadler}
\affiliation{Julius-Maximilians-Universit\"{a}t W\"{u}rzburg, Institut f\"{u}r Theoretische Physik und Astrophysik, Lehrstuhl f\"{u}r Astronomie, Emil-Fischer-Stra{\ss}e 31, 97074 W\"{u}rzburg, Germany}

\author{Dana Kirchner}
\affiliation{Julius-Maximilians-Universit\"{a}t W\"{u}rzburg, Institut f\"{u}r Theoretische Physik und Astrophysik, Lehrstuhl f\"{u}r Astronomie, Emil-Fischer-Stra{\ss}e 31, 97074 W\"{u}rzburg, Germany}

\author[0000-0001-6757-3098]{Georgios F. Paraschos}
\affiliation{ Finnish Centre for Astronomy with ESO, University of Turku, 20014 Turku, Finland}
\affiliation{ Aalto University Mets$\ddot a$hovi Radio Observatory, Mets$\ddot a$hovintie 114, 02540 Kylm$\ddot a$l$\ddot a$, Finland}
\affiliation{ Max-Planck-Institut f$\ddot u$r Radioastronomie, Auf dem H$\ddot u$gel 69, D-53121 Bonn, Germany}


\author{Florian R\"{o}sch}
\affiliation{Max-Planck-Institut f\"{u}r Radioastronomie, Auf dem H\"{u}gel 69, D-53121 Bonn, Germany}
\affiliation{Julius-Maximilians-Universit\"{a}t W\"{u}rzburg, Institut f\"{u}r Theoretische Physik und Astrophysik, Lehrstuhl f\"{u}r Astronomie, Emil-Fischer-Stra{\ss}e 31, 97074 W\"{u}rzburg, Germany}

\author{Wladislaw Schulga}
\affiliation{Julius-Maximilians-Universit\"{a}t W\"{u}rzburg, Institut f\"{u}r Theoretische Physik und Astrophysik, Lehrstuhl f\"{u}r Astronomie, Emil-Fischer-Stra{\ss}e 31, 97074 W\"{u}rzburg, Germany}

\author[0000-0003-0685-3621]{Mark Gurwell}
\affiliation{Center for Astrophysics $|$ Harvard \& Smithsonian, 60 Garden Street, Cambridge MA 02138, USA}

\author[0000-0002-3490-146X]{Garrett Keating}
\affiliation{Center for Astrophysics $|$ Harvard \& Smithsonian, 60 Garden Street, Cambridge MA 02138, USA}

\author[0000-0002-1407-7944]{Ramprasad Rao}
\affiliation{Center for Astrophysics $|$ Harvard \& Smithsonian, 60 Garden Street, Cambridge MA 02138, USA}

\author[0000-0001-7327-5441]{Emmanouil Angelakis}
\affiliation{Section of Astrophysics, Astronomy \& Mechanics, Department of Physics, National and Kapodistrian University of Athens,
Panepistimiopolis Zografos 15784, Greece}

\author[0000-0002-4184-9372]{Alexander Kraus}
\affiliation{Max-Planck-Institut f\"{u}r Radioastronomie, Auf dem H\"{u}gel 69,
D-53121 Bonn, Germany}

\author{Beatriz Ag\'{i}s-Gonz\'{a}lez}
\affiliation{Institute of Astrophysics, Foundation for Research and Technology-Hellas, GR-71110 Heraklion, Greece}

\author{Dmitry Blinov}
\affiliation{Institute of Astrophysics, Foundation for Research and Technology-Hellas, GR-71110 Heraklion, Greece}
\affiliation{Department of Physics, University of Crete, 70013, Heraklion, Greece}

\author{Anastasia Glykopoulou}
\affiliation{Institute of Astrophysics, Foundation for Research and Technology-Hellas, GR-71110 Heraklion, Greece}
\affiliation{Department of Physics, University of Crete, 70013, Heraklion, Greece}

\author[0009-0007-1918-577X]{Sara Capecchiacci}
\affiliation{Institute of Astrophysics, Foundation for Research and Technology-Hellas, GR-71110 Heraklion, Greece}
\affiliation{Department of Physics, University of Crete, 70013, Heraklion, Greece}

\author[0009-0004-1671-5454]{Alberto Floris}
\affiliation{Institute of Astrophysics, Foundation for Research and Technology-Hellas, GR-71110 Heraklion, Greece}
\affiliation{Department of Physics, University of Crete, 70013, Heraklion, Greece}
\affiliation{Istituto Nazionale di Astrofisica (INAF), Osservatorio Astronomico di Padova, IT-35122 Padova, Italy.}

\author{Panagiotis Fotis}
\affiliation{Institute of Astrophysics, Foundation for Research and Technology-Hellas, GR-71110 Heraklion, Greece}
\affiliation{Department of Physics, University of Crete, 70013, Heraklion, Greece}

\author{Alkmini Koukoutsi}
\affiliation{Institute of Astrophysics, Foundation for Research and Technology-Hellas, GR-71110 Heraklion, Greece}
\affiliation{Department of Physics, University of Crete, 70013, Heraklion, Greece}

\author{John A. Kypriotakis}
\affiliation{Institute of Astrophysics, Foundation for Research and Technology-Hellas, GR-71110 Heraklion, Greece}

\author[0009-0003-9365-9073]{Dimitrios A. Langis}
\affiliation{Institute of Astrophysics, Foundation for Research and Technology-Hellas, GR-71110 Heraklion, Greece}
\affiliation{Department of Physics, University of Crete, 70013, Heraklion, Greece}

\author{Dimitrios Rompogiannakis}
\affiliation{Institute of Astrophysics, Foundation for Research and Technology-Hellas, GR-71110 Heraklion, Greece}
\affiliation{Department of Physics, University of Crete, 70013, Heraklion, Greece}

\author[0009-0005-7962-6296]{Aristeidis Polychronakis}
\affiliation{Department of Physics, University of Crete, 70013, Heraklion, Greece}

\author{Vasiliki Tsioupli}
\affiliation{Institute of Astrophysics, Foundation for Research and Technology-Hellas, GR-71110 Heraklion, Greece}
\affiliation{Department of Physics, University of Crete, 70013, Heraklion, Greece}

     \author{Stavros Vogiatzis}
\affiliation{Institute of Astrophysics, Foundation for Research and Technology-Hellas, GR-71110 Heraklion, Greece}
\affiliation{Department of Physics, University of Crete, 70013, Heraklion, Greece}

\author{Orestis Zoumpoulakis}
\affiliation{Institute of Astrophysics, Foundation for Research and Technology-Hellas, GR-71110 Heraklion, Greece}
\affiliation{Department of Physics, University of Crete, 70013, Heraklion, Greece}

\author{Sumie Tochihara}
\affiliation{Department of Physics, Graduate School of Advanced Science and Engineering, Hiroshima University Kagamiyama, 1-3-1 Higashi-Hiroshima, Hiroshima 739-8526, Japan}

\author{Ryo Imazawa}
\affiliation{Department of Physics, Graduate School of Advanced Science and Engineering, Hiroshima University Kagamiyama, 1-3-1 Higashi-Hiroshima, Hiroshima 739-8526, Japan}

\author{Mahito Sasada}
\affiliation{Department of Physics, Tokyo Institute of Technology, 2-12-1 Ookayama, Meguro-ku, Tokyo 152-8551, Japan}

\author{Yasushi Fukazawa}
\affiliation{Department of Physics, Graduate School of Advanced Science and Engineering, Hiroshima University Kagamiyama, 1-3-1 Higashi-Hiroshima, Hiroshima 739-8526, Japan}
\affiliation{Hiroshima Astrophysical Science Center, Hiroshima University 1-3-1 Kagamiyama, Higashi-Hiroshima, Hiroshima 739-8526, Japan}
\affiliation{Core Research for Energetic Universe (Core-U), Hiroshima University, 1-3-1 Kagamiyama, Higashi-Hiroshima, Hiroshima 739-8526, Japan}

\author{Koji S. Kawabata}
\affiliation{Department of Physics, Graduate School of Advanced Science and Engineering, Hiroshima University Kagamiyama, 1-3-1 Higashi-Hiroshima, Hiroshima 739-8526, Japan}
\affiliation{Hiroshima Astrophysical Science Center, Hiroshima University 1-3-1 Kagamiyama, Higashi-Hiroshima, Hiroshima 739-8526, Japan}
\affiliation{Core Research for Energetic Universe (Core-U), Hiroshima University, 1-3-1 Kagamiyama, Higashi-Hiroshima, Hiroshima 739-8526, Japan}

\author{Makoto Uemura}
\affiliation{Department of Physics, Graduate School of Advanced Science and Engineering, Hiroshima University Kagamiyama, 1-3-1 Higashi-Hiroshima, Hiroshima 739-8526, Japan}
\affiliation{Hiroshima Astrophysical Science Center, Hiroshima University 1-3-1 Kagamiyama, Higashi-Hiroshima, Hiroshima 739-8526, Japan}
\affiliation{Core Research for Energetic Universe (Core-U), Hiroshima University, 1-3-1 Kagamiyama, Higashi-Hiroshima, Hiroshima 739-8526, Japan}

\author[0000-0001-7263-0296]{Tsunefumi Mizuno}
\affiliation{Hiroshima Astrophysical Science Center, Hiroshima University, 1-3-1 Kagamiyama, Higashi-Hiroshima, Hiroshima 739-8526, Japan}

\author{Tatsuya Nakaoka}
\affiliation{Hiroshima Astrophysical Science Center, Hiroshima University 1-3-1 Kagamiyama, Higashi-Hiroshima, Hiroshima 739-8526, Japan}

\author{Hiroshi Akitaya}
\affiliation{Planetary Exploration Research Center, Chiba Institute of Technology 2-17-1 Tsudanuma, Narashino, Chiba 275-0016, Japan}

\author{Takahiro Akai}
\affiliation{Department of Physics, Graduate School of Advanced Science and Engineering, Hiroshima University Kagamiyama, 1-3-1  Higashi-Hiroshima, Hiroshima 739-8526, Japan}

\begin{abstract}

We report on a nine-month monitoring campaign of the blazar PG\,1553+113, relying on three observations carried out in 2025 with the Imaging X-ray Polarimetry Explorer ({\it IXPE}) and supported by multi-wavelength facilities. The source displayed pronounced variability across the electromagnetic spectrum, with X-ray flux changes by up to a factor of $\sim5$ and complex evolution of the optical polarization properties, including one of the largest (exceeding $150^{\circ}$) and fastest rotations in the electric vector position angle (EVPA) ever recorded. This swing of the EVPA was also accompanied by a temporary drop of the optical polarization degree to nearly zero. Significant X-ray polarization was observed during the third {\it IXPE} pointing, with a polarization degree $\Pi_{\rm X}\,=(\,18.4\,\pm\,5.8)\%$ and  $\Psi_{\rm X}\,=\,74^{\circ} \pm 9^{\circ}$ in the 2--8~keV band, while only upper limits were obtained in the first two epochs. The optical data show that the second {\it IXPE} observation occurred shortly after a dramatic optical polarization event characterized by a rapid EVPA swing and strong depolarization. Two possible scenarios may explain the broadband polarimetric behavior: (i) the superposition of two emitting regions with nearly orthogonal magnetic field configurations and variable relative contributions, and (ii) the interaction of a single emitting region with a shock that temporarily reorders the magnetic field. In both cases, the data support a picture in which the X-ray and optical emissions arise from closely related but not strictly co-spatial regions within a dynamically evolving, magnetically structured jet.

\end{abstract}


\section{introduction}

Blazars are a subclass of active galactic nuclei (AGN) characterized by relativistic jets of particles oriented almost parallel to our line of sight \citep[e.g.,][]{Hovatta2015, Blandford2019}. This results in highly variable emission observable at all frequencies of the electromagnetic domain \citep[e.g.,][]{Raiteri2025}. Based on their spectral energy distribution (SED), which is typically double humped, blazars are classified into three categories depending on the frequency ($\nu_{\rm peak}$) of the peak of the low-energy hump: low-synchrotron-peak (LSP), intermediate-synchrotron-peak (ISP), and high-synchrotron-peak (HSP) objects. The low frequency peak is ascribed to synchrotron emission, and for these different sub-classes it is generally observed in the infrared, optical, and X-ray bands, respectively \citep{Fossati1998, Abdo2010}.

For decades, the physics of the jets has been studied using multi-wavelength data from space observatories, including {\it Fermi} Large Area Telescope \citep[LAT,][]{Atwood2009}, the Neil Gehrels Swift Observatory \citep[hereafter \textit{Swift};][]{Gehrels2004}, {\it XMM-Newton}, and ground-based radio and optical telescopes  \citep[e.g.,][]{Mead1990, Carini1992,Piconcelli2005,Marscher2008,Ajello2009,Smith2009,Ackermann2015,Acciari2020,Ajello2020}. The Imaging X-ray Polarimetry Explorer \citep[{\it IXPE;\rm}][]{Weisskopf2022} has revolutionized this field, allowing for direct measurements of X-ray polarization. This provides unique constraints on acceleration mechanisms of the particle population responsible for the observed broadband emission, and unprecedented insights on the structure of the magnetic field within the X-ray emitting region of the jet \citep[e.g.,][]{Liodakis2022,DiGesu2022-Mrk421,DiGesu2023,Middei2022,Peirson2023,Kouch2024,Agudo2025}; see  \citet[]{Marscher2024} for a summary. 

The large number of {\it IXPE} observations, accompanied by multi-frequency polarimetric campaigns, have also enabled ensemble analyses: for LSP objects the observations \citep[e.g.,][]{Marshall2024,Agudo2025} suggest that the particle populations giving rise to the keV emission of these sources are likely of leptonic origin \citep[but see also][]{Tavecchio2025}. As noted by \citet{Capecchiacci2025}, in all the HSP sources observed by {\it IXPE}, the polarization degree in X-rays is systematically higher than at optical and mm--radio wavelengths. This finding is compatible with jets being energy stratified, with particles being accelerated in shocks or some other localized regions, from which they propagate while losing energy \citep[e.g.,][]{Liodakis2022,Marscher2024}.{\it IXPE} has also performed multi-epoch observational campaigns of HSP blazars such as Mrk\,501 \citep{Chen2024}, which maintained fairly constant polarization during 2022 and 2023, and Mrk\,421 \citep{Kim2024}, which on the other hand displayed dramatic changes in the X-ray polarization degree. 

Here we report on novel broadband spectro-polarimetric observations of PG\,1553+113 (RA=15h 55m 43.0440s, Dec$=+11^{\circ}$ $11'$ $24.365''$, J2000), a HSP blazar ($\nu_{\rm peak}\approx3.9\times10^{15}$; \citealp{Ajello2020}). This source, a known TeV emitter \citep{Aleksic2012} likely located at $z=0.433$ \citep[e.g.,][]{Nicastro2018}, is one of the few blazars exhibiting quasi-periodic variability patterns \citep[e.g.,][]{Penil2020,Abdollahi2024}. Specifically, it displays a compelling $\sim$2.2-year quasi-periodicity in its $\gamma$-ray flux and in the optical/UV bands \citep{Ackermann2015, Tavani2018, MAGIC2024}.
At X-ray energies, the situation is more complex and still debated. While some studies report no statistically significant periodicity 
\citep[e.g.,][]{Aniello2024, MAGIC2024}, others have found tentative hints 
of a similar quasi-periodic pattern \citep{Penil2026} and interpreted X-ray flares within the framework of jet precession in a binary supermassive black hole (SMBH) system \citep{Huang2021}. Therefore, while a long-term modulation may exist across the electromagnetic spectrum, it is most likely masked in X-rays by pronounced stochastic variability.
The origin of this quasi-periodic modulation is still 
poorly understood. 
The most widely discussed scenario involves a binary SMBH system at sub-parsec scales \citep[e.g.,][]{Ackermann2015, Tavani2018, Abdollahi2024}, where the observed $\sim$2.2-year periodicity could arise from jet precession, disk perturbations, or periodic modulation of the accretion rate.  In this framework, the reported hints of a longer $\sim$22-year 
timescale \citep{Adhikari2024} may indicate a more complex, hierarchical variability, possibly associated with large-scale precession of the accretion disk or orbital evolution. However, alternative interpretations do not necessarily require a binary system. These include geometric effects within a single SMBH scenario, such as jet precession driven by instabilities, or variations in the viewing angle, as well as intrinsic processes within the jet. Supporting a geometric origin, recent analyses have shown that the $\sim$2.2-year quasi-periodicity in $\gamma$-rays is nearly achromatic \citep{Madero2026}. This behavior is naturally explained by periodic 
changes in the Doppler boosting factor, rather than by intrinsic variations in the particle population, thus favoring scenarios involving jet orientation effects.

Given its unique observational properties, PG\,1553+113 is a prime target for X-ray polarization measurements by {\it IXPE}. \citet{Middei2023} have already carried out {\it IXPE} and multi-wavelength observations, reporting on the blazar's broadband polarimetric properties. The X-ray polarization degree was observed to be  $\Pi_{\rm X}=(10\pm 2)\%$ in the 2--8 keV range, with an Electric Vector Polarization Angle (EVPA) of $\Psi_{\rm X}=(86^\circ \pm 8^\circ$). Interestingly, the optical EVPA exhibited a large swing of approximately \(125^\circ\) over a time span of days, at a rate of \(\sim 17^\circ\) per day, while no corresponding variation was detected in the X-ray or radio bands. This implies that the X-ray and optical emission regions are not co-spatial, since they have distinct magnetic field geometries. 

During 2025, we again monitored PG\,1553+113 with {\it IXPE} and coordinated ground- and space-based facilities. In this work, we report on the broadband polarimetric properties collected within this program. In \S2 we describe the data reduction procedures and present the data, while \S3 reports on the spectro-polarimetric analyses. In \S4 we report on the multi-wavelength polarimetric data, while in \S5 we comment the results and draw our conclusions.

\section{Data reduction}
PG\,1553+113 was the target of an intense monitoring campaign during 2025 involving various polarimeters and CCD detectors. In the X-ray band we observed PG\,1553+113 with {\it IXPE}, \textit{XMM-Newton} \citep[][]{Jansen2001} and {\it Swift}. In Table~\ref{obslog} we list the related X-ray observations for {\it IXPE} and {\it XMM-Newton}, while for {\it Swift} a similar table is shown in Appendix A.

\begin{table}[ht]
\centering
\caption{Log of the {\it IXPE}, {\it XMM} observations of PG\,1553+113 used in this work.}
\label{obslog}
\begin{tabular}{lcc}
\hline\hline
 ObsID & Start date (UTC) & Exposure (ks) \\
\hline
\multicolumn{3}{c}{{\it IXPE}} \\
\hline
04001701 & 2025-02-01 14:50:48 & 119.6 \\
04001801 & 2025-07-06 11:57:54 & 117.9 \\
04001901 & 2025-09-11 07:47:39 & 115.0 \\
\hline
\multicolumn{3}{c}{{\it XMM--Newton}} \\
\hline
0920901101 & 2025-02-02 09:15:05 & 13.6 \\
0920902301 & 2025-07-28 15:33:14 & 7.0 \\
0920902501 & 2025-08-26 18:39:39 & 15.5 \\
\hline
\end{tabular}
\end{table}

\subsection{{\it IXPE} data reduction}

We processed the {\it IXPE} level-1\footnote{Detector Unit 2 (DU2) has been operating in an anomalous manner since 14 April 2025; hence, for our observations after 10 May 2025, the DU2 event files were not available in time to include in this study.} event files available for download from the \textit{HEASARC} archive. The Stokes $I$, $Q$, and $U$ spectra were derived independently for the available detector units (DUs) through use of the \textsc{ixpeobssim} package (version 31.1.1; \citealt{pesce2019,baldini2022}), applying the background filtering strategy described in \citet{DiMarco2023}. In addition, the spectra were generated by utilizing the weighted analysis technique introduced by \citet[][]{dimarco2022}, corresponding to the STOKES = NEFF option in XSELECT. Source events were selected from a circular extraction region with a radius of 0.95$'$ centered on the target, whereas the background was estimated from an annular region with inner and outer radii of 1.2$'$ and 3.5$'$, respectively. This selection has been demonstrated to provide optimal polarization sensitivity \citep{DiMarco2023}. The $I$-Stokes spectra were rebinned to ensure a minimum signal-to-noise ratio of 7 per energy channel, while a fixed energy bin width of 280 eV was adopted for the $Q$ and $U$ spectra.

\subsection{XMM-Newton Data Reduction}
PG\,1553+113 was also observed three times by {\it XMM-Newton} as close in time as possible with the {\it IXPE} pointings. Our analysis focused on data collected with the {\it EPIC}-pn instrument \citep{Struder2001}, operated in Small Window mode with the medium optical blocking filter. The data reduction was carried out using the standard {\it XMM-Newton} Science Analysis System (SAS version 230412; \citealt{Gabriel2004}). The selection of source extraction regions and identification of time intervals affected by enhanced background were performed iteratively, following the procedure described by \citet{Piconcelli2004}, with the aim of optimizing the signal-to-noise ratio. The background spectra were extracted from circular regions with a radius of 50$''$. The final, third-level products were rebinned to include a minimum of 30 counts per channel and to ensure that the instrumental energy resolution was not oversampled by more than a factor of three. The observed net count rate remained well below the threshold above which pile-up effects become significant for EPIC-pn Small Window observations \citep[e.g.,][]{Jethwa2015}. A dedicated pile-up assessment was also carried out using the SAS task \emph{epatplot}, confirming that pile-up effects were negligible.

\subsection{Swift (XRT and UVOT)}

{\it Swift} observed PG\,1553+113 both prior to and following the {\it IXPE} observations, and throughout 2025. A total of 48 individual pointings, each with an exposure time of $\sim$1 ks, were acquired in either Photon Counting (PC) or Windowed Timing (WT) mode, depending on the flux of the source. The data were processed and calibrated using an automated pipeline that relies on the standard {\it XRTDAS} software package\footnote{\url{https://sda2006.ts.infn.it/presentazioni/capalbi.pdf}}. In order to extract the source events, the pipeline handles observations executed in the two observing modes using a circular/annular region centered on the target, depending on the presence of pile-up \citep[see][for details]{Middei2022a}. The background contribution was estimated from a concentric annular region with inner and outer radii of $120''$ and $150''$ for observations in PC, and with synthetic backgrounds for the WT observations. The resulting spectra were rebinned to ensure at least 25 counts per energy bin, allowing for the application of $\chi^2$ statistics in the subsequent spectral fitting. 
The Ultraviolet/Optical Telescope (UVOT) data were processed with {\tt uvotimsum} to co-add exposures, and with {\tt uvotsource} to extract photometry in all available filters ($V$, $B$, $U$, $UVW1$, $UVM2$, and/or $UVW2$). The observed UVOT fluxes were then used to build broadband SEDs and to monitor optical/UV variability concurrent with X-ray polarimetric measurements.
In Table~\ref{Swift} we report the relevant information for the {\it Swift} exposures used in this work.

\section{Data analysis}

Joint X-ray spectral and polarimetric modeling was carried out with the \textsc{XSPEC} package \citep{Arnaud1996}, adopting the weighted analysis approach described by \citet{dimarco2022}. The effects of Galactic photoelectric absorption were included in the fits through the {\tt TBabs} model, with the hydrogen column density fixed to $N_{\rm H}$ = $3.62 \times 10^{20}$ cm$^{-2}$, as derived from the HI4PI survey \citep{HI4PI2016}. Whenever a distance-dependent quantity for PG\,1553+113 was required, we assumed a standard cosmological model characterized by $H_0 = 70$ km s$^{-1}$ Mpc$^{-1}$ and $\Lambda_0 = 0.73$. Finally, all the errors quoted in the tables correspond to 68\% uncertainties, except for upper/lower limits, which are at the 99.7\% confidence level.

\subsection{Broadband properties of PG\,1553+113 during 2025}
As described in \S2.3, PG\,1553+113 was observed 48 times with {\it Swift} between early January and late September 2025. Each spectrum was fitted with a phenomenological model, a logarithmic parabola \citep[e.g.,][]{Massaro2004}, absorbed by gas in the Milky Way. The fits provide the spectral shape ($\alpha$), curvature parameter ($\beta$), and flux normalization for a selected pivot energy of 1 keV \citep[e.g.,][]{Giommi2021}. This model adequately accounts for all spectra; the best-fit parameters are reported in Table~\ref{Swift}.

\begin{figure}
    \centering
    \includegraphics[width=0.99\columnwidth]{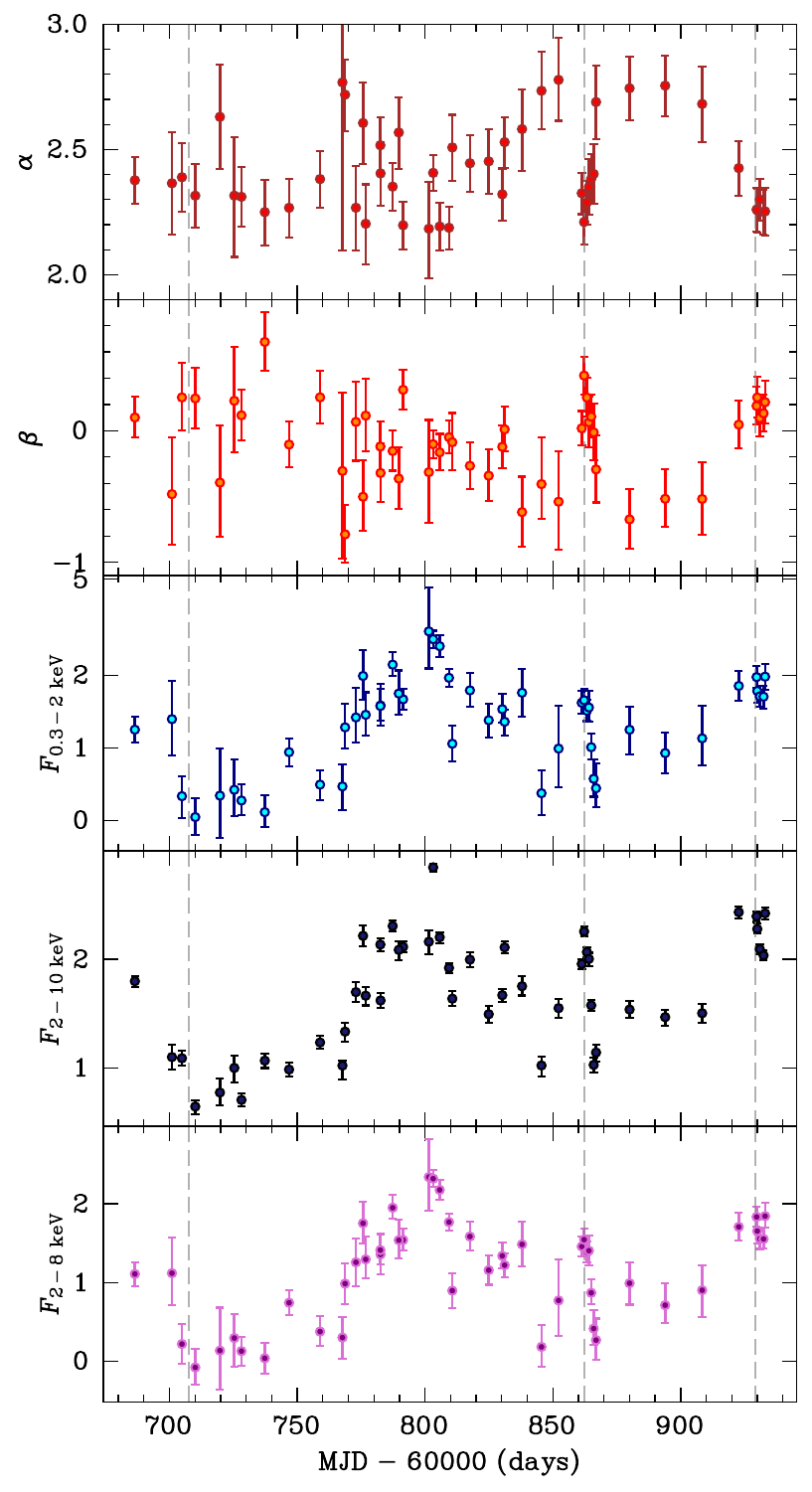}
    \caption{Temporal evolution of the spectral properties ($\alpha$ and $\beta$ panels) and the flux in multiple bands of PG\,1553+113. Vertical dashed lines identify the epochs of the \textit{IXPE} exposures. Fluxes are in units of 10$^{-11}$ erg cm$^{-2}$ s$^{-1}$.}
    \label{swiftlc}
\end{figure}

\begin{figure*}
    \centering
    \includegraphics[width=0.99\textwidth]{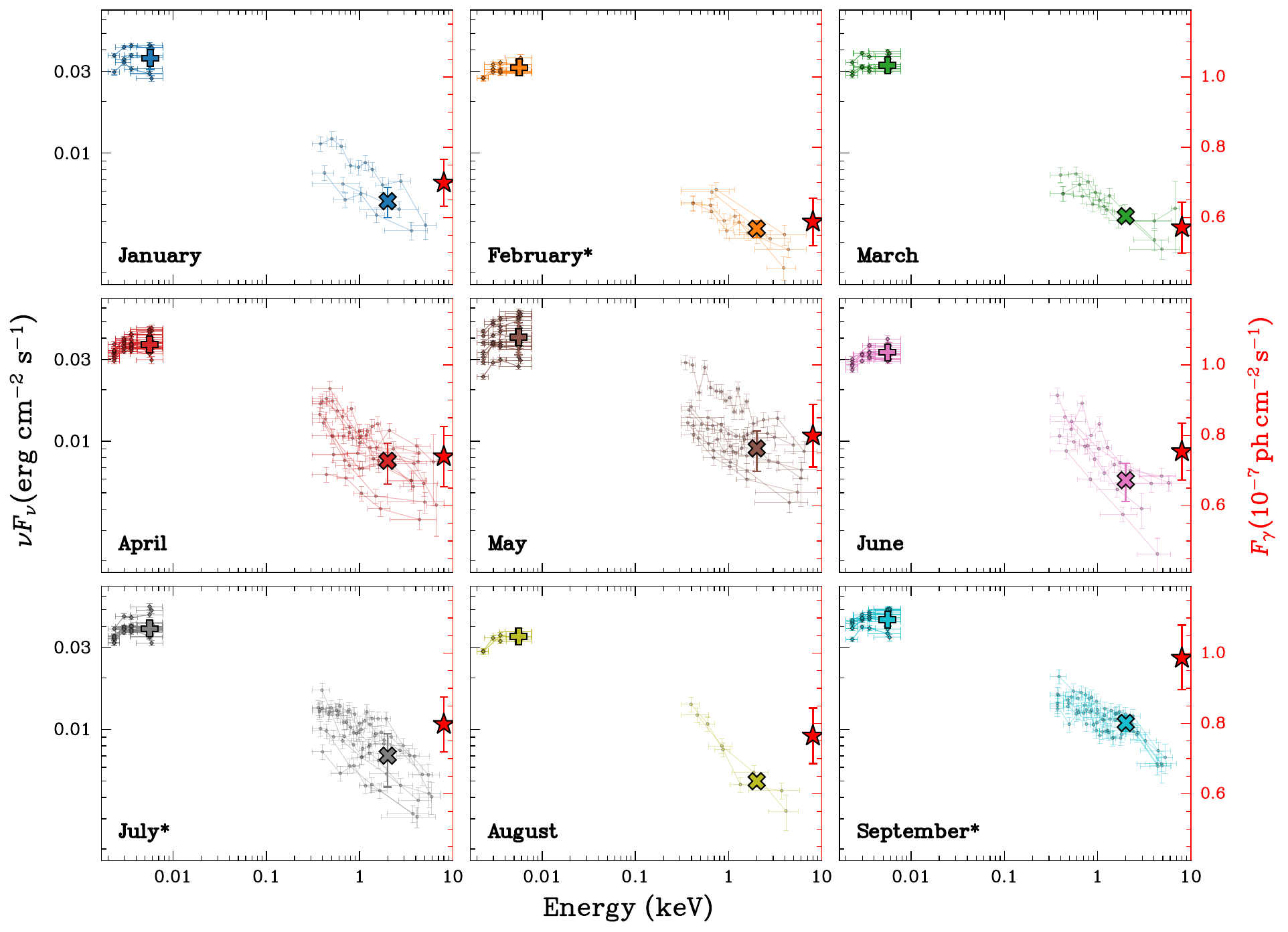}
    \caption{Monthly SEDs of PG\,1553+113 during January--September 2025, covering the optical/UV and X-ray bands. Each panel shows the collected {\it Swift} UVOT and XRT data. Both the ultraviolet and X-ray emissions exhibit significant variability in amplitude and spectral shape. The black asterisks  near the name of months in the bottom of panels mark the months during which {\it IXPE} observed the source. The corresponding monthly {\it Fermi}-LAT flux level \citep[][]{Abdollahi2023} is shown as a red star in each sub-panel. The right y-axis provides the scale for the $\gamma$-ray flux.}
    \label{SEDfig}
\end{figure*}

During the analyzed period, PG\,1553+113 shows significant flux variability, by up to a factor of $\sim$5 in the 2--8 keV band. The spectral shape and curvature also vary substantially on both daily and monthly timescales. From Fig.~\ref{swiftlc}, we find that {\it IXPE} observed PG\,1553+113 at three different flux levels within the range $F\sim(0.5-2)\times 10^{-11}$ erg cm$^{-2}$ s$^{-1}$. The XRT spectra and UVOT data can be used to build the broadband SED of PG\,1553+113, which we show in Fig.~\ref{SEDfig}.
In this figure, we present the temporal evolution of the source SEDs and compare them with the corresponding $\gamma$-ray flux measured by {\it Fermi}-LAT \citep[][]{Abdollahi2023}. Significant variability is observed in both amplitude and spectral shape across the optical/UV and X-ray bands, on timescales ranging from weeks to months.\\
To further investigate the multi-wavelength variability and search for correlations between different energy bands, we have constructed monthly averaged SEDs.
For the X-ray band, we estimated the monochromatic flux at a reference energy of 2 keV ($\nu F_{\nu}$ at 2 keV). This flux was derived for each observation through a log--log linear interpolation between the two nearest energy bins. For the UV band, we adopted the UVW1 filter \citep[$\lambda_{\mathrm{eff}} \approx 2486\,\text{\AA}$,][]{Breeveld2011} as a proxy for the ultraviolet emission.
The monthly mean fluxes ($\langle F \rangle$) were computed by averaging observations within each calendar month. The associated uncertainties were estimated as the standard deviations of the corresponding fluxes. When only a single observation was available, the corresponding instrumental uncertainty was adopted.
We then performed correlations between the $\gamma$-ray integrated {\it Fermi}-LAT flux and the X-ray and UV fluxes, finding strong correlations with Pearson coefficients of $R = 0.890$ ($p = 1.3 \times 10^{-3}$) for the X-ray versus $\gamma$-ray flux, and $R = 0.887$ ($p = 1.4 \times 10^{-3}$) for the UV versus $\gamma$-ray flux.

\subsection{{\it XMM-Newton} spectral analysis}
Although the {\it XMM-Newton} observations are not fully simultaneous with the {\it IXPE} pointings, the large effective area in the soft band can help constrain the overall spectral shape of the source. For this reason, we have modeled the 3 {\it EPIC}-pn spectra with the simple phenomenological model applied above for the XRT snapshots. The fitting procedure is the same, and we report the corresponding results in Table~\ref{Istokes}. The {\it XMM-Newton} spectra are compatible with a softening of the spectrum when the source becomes brighter (softer when brighter). During these epochs, no strong curvature is observed in the spectra. 

\begin{table}[h!]
\centering
\caption{Best-fit parameters of the spectral fits to the{\it XMM-Newton} data.}
\begin{tabular}{lccc}
\hline
\textbf{comp. \& param.} & \textbf{Obs 1} & \textbf{Obs 2} & \textbf{Obs 3} \\
\hline
\texttt{logpar $\alpha$} & $2.61 \pm 0.01$ & $2.67 \pm 0.01$ & $2.68 \pm 0.01$ \\
\texttt{logpar $\beta$} & $0.04 \pm 0.03$ & $-0.02 \pm 0.03$ & $0.03 \pm 0.02$ \\
\texttt{norm} ($\times 10^{-3}$) & $5.87 \pm 0.03$ & $6.81 \pm 0.05$ & $8.59 \pm 0.03$ \\
\hline
\textbf{$F_{\rm 0.3-2\,keV}\,(\times 10^{-11}$)} & $1.65\pm0.01$ & $1.97\pm0.01$ & $2.48\pm0.01$ \\
\textbf{$F_{\rm 2-10\,keV}\,(\times 10^{-12}$)} & $6.03\pm0.12 $ & $6.85\pm0.20$ & $8.09\pm0.11$ \\
\hline
\textbf{$\chi^2$/d.o.f.} & $149 / 136$ & $130 / 114$ & $186 / 152$ \\
\hline\label{Istokes}
\end{tabular}
\end{table}

\subsection{{\it IXPE} spectro-polarimetric analysis}
\begin{figure*}
    \centering
    \includegraphics[width=0.98\textwidth]{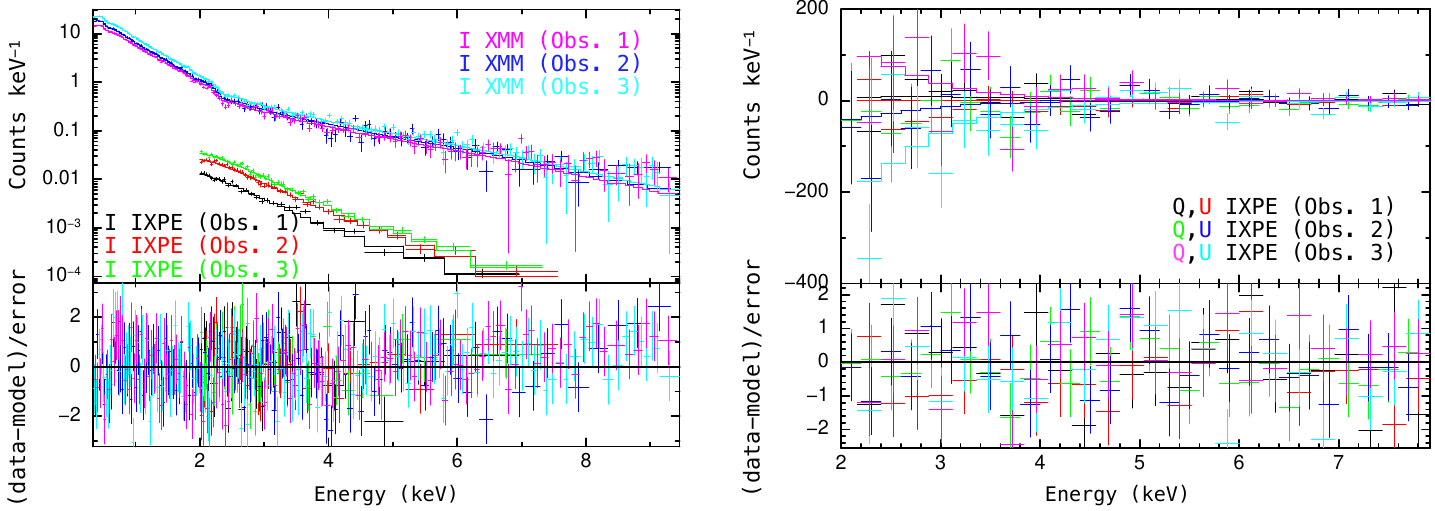}
    \caption{{\it Left panel}: Best fit to the $I$ Stokes (i.e., total flux density) spectra derived for the \textit{IXPE} and {\it XMM-Newton} observations reported in Table~\ref{obslog}. {\it Right panel}: The corresponding best-fit to the $Q$ and $U$ Stokes spectra.}
    \label{swiftbestfit}
\end{figure*}

\begin{figure}
    \centering
    \includegraphics[width=0.98\columnwidth]{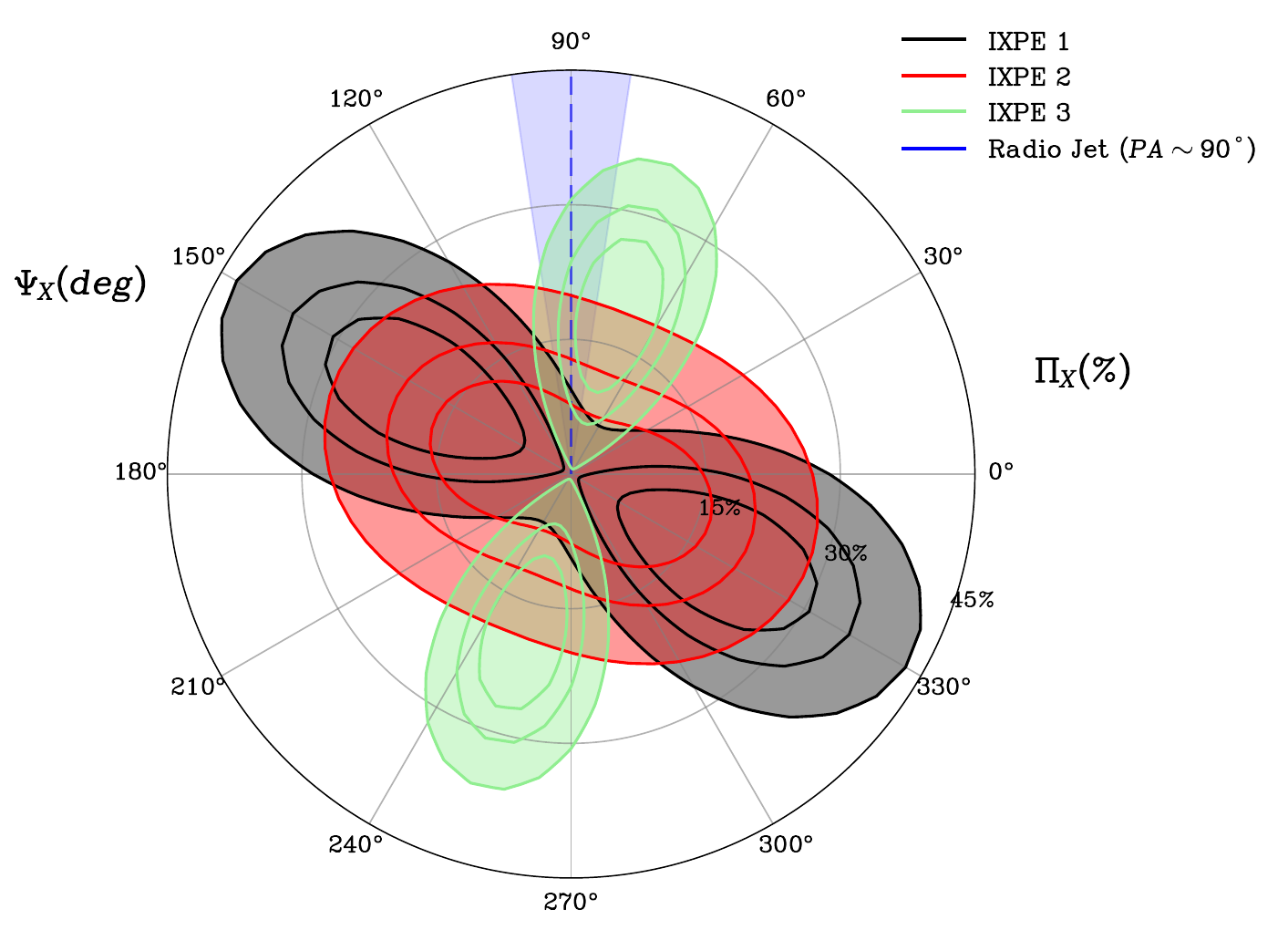}
    \caption{Confidence regions for the polarization properties of PG\,1553+113 at 68\%, 90\% and 99\% confidence level. Only Obs. 3 enables us to determine both the X-ray polarization degree and angle. The jet projection is taken from \citet[][]{Capecchiacci2025}.}
    \label{cont}
\end{figure}

The strong short-term variability displayed by PG\,1553+113 prevents us from performing joint {\it IXPE--XMM} spectral fits. Thus, we fit the three sets of {\it IXPE} Stokes-parameter spectra, adopting the \texttt{TBabs*constant*polconst*logpar}  model for each distinct exposure in \textsc{XSPEC}. The \texttt{constant} parameter takes into account the cross-calibration among the different DUs, \texttt{logpar} models the non-thermal continuum of the source (in this case we set the pivot energy to 3 keV, within the {\it IXPE} operating band), while \texttt{polconst}, operating on the $Q$ and $U$ spectra, returns the degree $\Pi_{\rm X}$ and position angle $\Psi_{\rm X}$ of the 2--8 keV polarization. The fit to the data is straightforward (see Fig.~\ref{swiftbestfit}) and the corresponding results are listed in Table\,~\ref{ixperesults}. Only for {\it IXPE} observation 3 are the polarization properties of PG\,1553+113 constrained, and for this measurement we obtain values of $\Pi_{\rm X}=(18.4 \pm 5.8)\%$ and
$\Psi_{\rm X}=74^\circ \pm 9^\circ$. Only upper limits are found for observations 1 and 2, thus in Table\,~\ref{ixperesults} we report their corresponding values at the 3$\sigma$ confidence level. Interestingly, if we considered only $1\sigma$ uncertainties, for Obs.\ 1 we would obtain $\Pi_{\rm X}=(18.4\pm8.1)\%$ and $\Psi_{\rm X}=150^\circ\pm13^\circ$, while for Obs.\ 2 we would obtain $\Pi_{\rm X}\,<\,$12\%. In Fig.~\ref{swiftbestfit} we show the best fit to the $I$ Stokes spectra for both {\it XMM-Newton} and {\it IXPE} data, as well as the best-fit to the $Q$ and $U$ Stokes spectra, while the corresponding contours for the polarization properties are displayed in Fig.~\ref{cont}. 

To further search for the presence of any polarimetric signal in the {\it IXPE} data, we also fit the three exposures jointly. This only returns an upper limit to the polarization degree $\Pi_{\rm X}<13\%$, this fit being slightly worse than that quoted in Table~\ref{ixperesults}. As an additional check, we assume the polarization degree to be constant across the observations and allow the polarization angle to vary. This leads to a fit with a statistic compatible with our best fit and a ``constant'' $\Pi_{\rm X}\,=(\,14.6\pm3.9)\%$, with polarization angles of $\Psi_{\rm X}\,>110^\circ$ and $\Psi_{\rm X}\,=74^\circ\pm10^\circ$ for Obs.\ 1 and Obs. 3, respectively, while this quantity is unconstrained by the data in Obs.\ 2.

\begin{table}[h!]
\centering
\caption{Best-fit model for the 3 {\it IXPE} observations}\label{ixperesults}
\begin{tabular}{lccc}
\hline
\textbf{comp. \& param.} & \textbf{Obs 1} & \textbf{Obs 2} & \textbf{Obs 3} \\
\texttt{logpar $\alpha$} 
& $2.59 \pm 0.07$ 
& $2.84 \pm 0.24$ 
& $2.70 \pm 0.05$ \\

\texttt{logpar $\beta$} 
& $-0.77 \pm 0.56$ 
& $1.01 \pm 0.90$ 
& $0.18 \pm 0.40$ \\

\texttt{norm} ($\times 10^{-4}$) 
& $2.27 \pm 0.08$ 
& $4.92 \pm 0.16$ 
& $6.54 \pm 0.15$ \\
\texttt{polconst $\Pi_{\rm X}$}
& $<39\%$ 
& $<24\%$  
& $18.4 \pm 5.8$ \\
\texttt{polconst $\Psi_{\rm X}$} 
& $--$ 
& $--$ 
& $74^\circ \pm 9^\circ$ \\
\texttt{DU2/DU1} 
& $0.97 \pm 0.04$ 
& $--$ 
& $--$ \\
\texttt{DU3/DU1} 
& $0.96 \pm 0.03$ 
& $0.89 \pm 0.02$ 
& $0.89 \pm 0.02$ \\
\hline
\texttt{$F_{\rm 2-8\,keV}\,(\times 10^{-12})$} 
& $4.2 \pm 0.3$ 
& $7.5 \pm 0.6$ 
& $10.8 \pm 0.2$ \\
\hline
\texttt{$\chi^2$/d.o.f.} 
& $205/236$ 
& $157/186$ 
& $226/192$ \\
\hline
\end{tabular}
\end{table}

\section{Multi-wavelength polarimetric data}

\begin{figure*}
    \centering
    \includegraphics[width=0.98\textwidth]{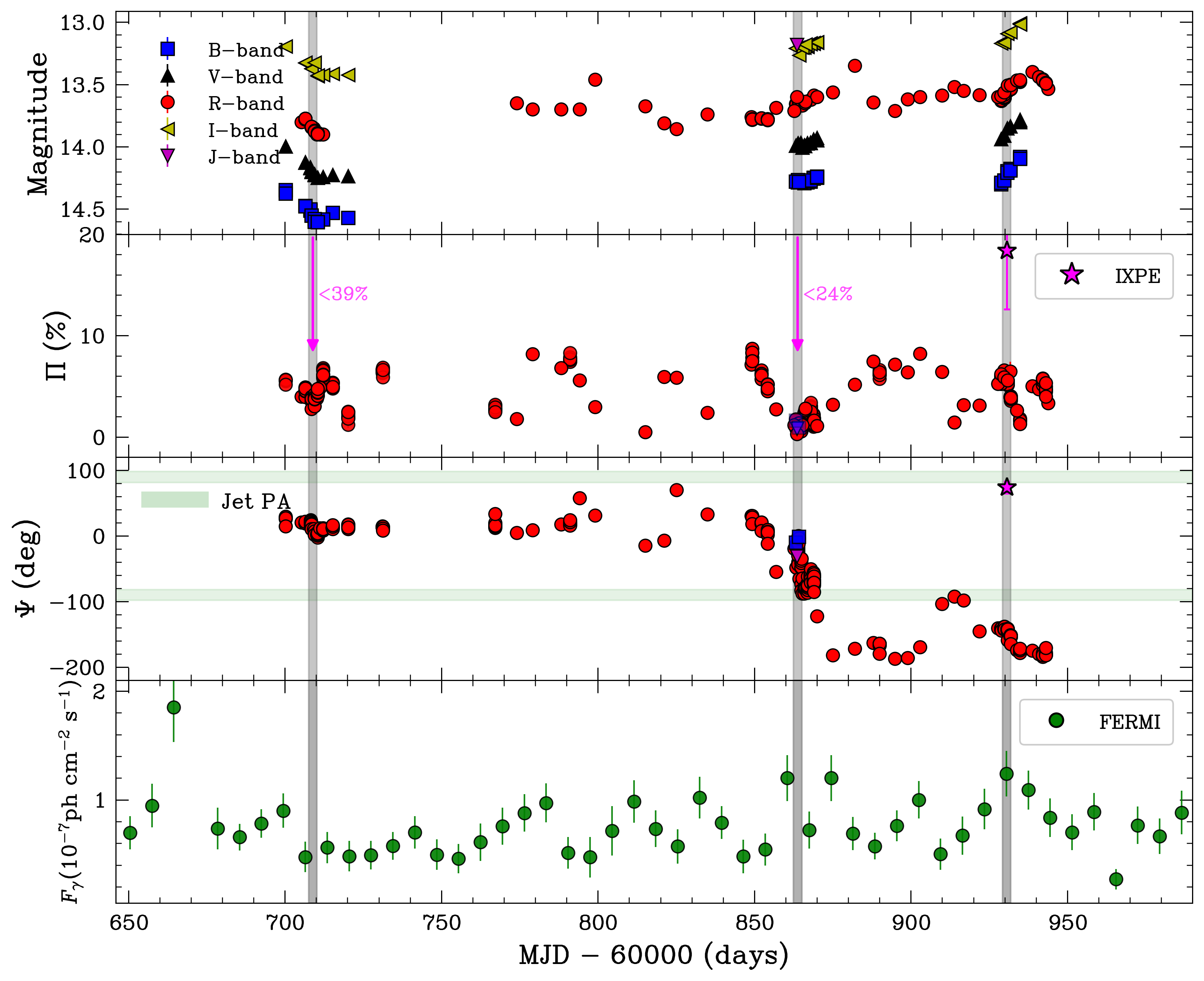}
    \caption{Temporal evolution of the flux (optical and $\gamma$-ray) and the optical polarization properties of PG\,1553+113. From top to bottom, the panels show the source brightness expressed in magnitudes; the optical polarization degree; the polarization angle; and the $\gamma$-ray light curve in the 0.1--100~GeV energy range, with the weekly photon flux measured by the \textit{Fermi}-LAT during the \textit{IXPE} monitoring campaign. The time intervals corresponding to the \textit{IXPE} observations are highlighted by gray shaded regions. Error bars indicate $1\sigma$ uncertainties. {\it IXPE} upper limits (arrows) and measurements (stars) are shown in magenta. Finally, the horizontal green bars shows the direction of the jet.}
    \label{mwl}
\end{figure*}

\begin{figure*}
    \centering
    \includegraphics[width=0.98\textwidth]{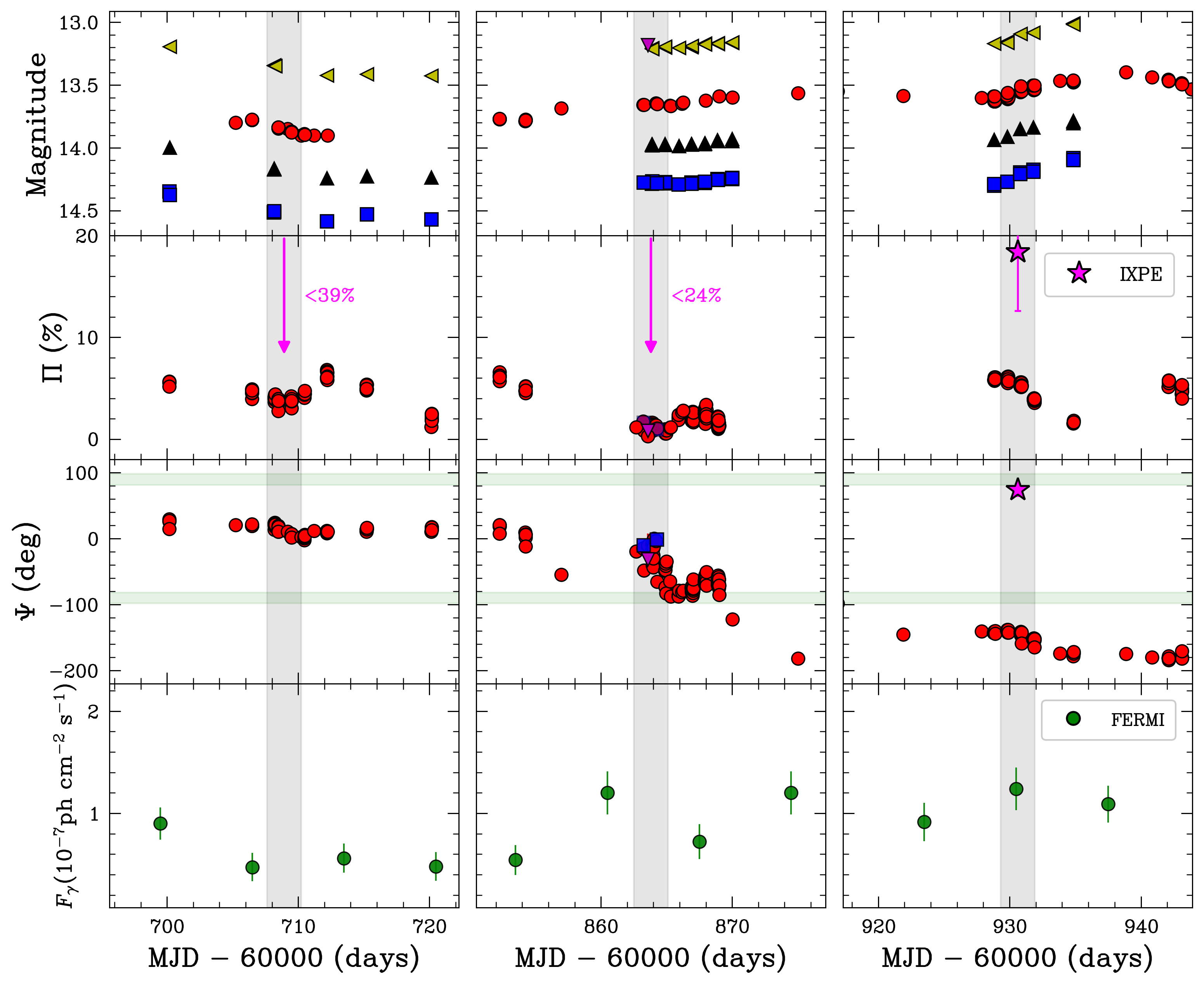}
    \caption{Zooms extracted from Fig.~\ref{mwl}. The different panels focus on the multi-wavelength behavior of PG\,1553+113 before and after the {\it IXPE} pointings. The quantities in the different sub-plots are the same as for Fig.~\ref{mwl}.}
    \label{mwlzoom}
\end{figure*}

As supplements to the {\it IXPE} multi-epoch observations, different ground-based observatories monitored PG\,1553+113 at both radio and optical wavelengths: Calar Alto Observatory \cite[CAFOS; ][]{juan_escudero:2023,escudero2024},  Effelsberg 100-m telescope \cite[QUIVER and TELAMON programs; ][]{Krauss2003,Myserlis2018,Eppel2024,Myserlis2025}, the Higashi-Hiroshima Observatory \cite[Kanata telescope - HONIR; ][]{kawabata_new_1999,akitaya_honir_2014}, the Nordic Optical Telescope \cite[ALFOSC; ][]{Nilsson2018,MAGIC2018}, the LX-200 telescope \citep[St. Petersburg University, Russia;][]{Larionov2008}, Boston University's Perkins Telescope \cite[PRISM camera;][]{Jorstad2010}, Skinakas observatory \cite[RoboPol; ][]{Ramaprakash2019,Blinov2021} and the Submillimeter Array \cite[SMAPOL program; ][]{Myserlis2025}. Details on the data analysis of the individual observatories and monitoring programs can be found in, e.g., \citet[][]{Liodakis2022,DiGesu2023,Peirson2023,Liodakis2025}. The emission from PG\,1553+113 does not contain significant host-galaxy contamination, hence no correction has been performed to the optical flux or polarization degree. In Figs.~\ref{mwl} and \ref{mwlzoom}, we show how the optical magnitudes, as well as the optical polarization degree and angle, of this blazar vary as a function of time. 

Until MJD 60849, the optical R-band polarization fluctuated about an average value of $\langle\Pi_{\rm R\,band}\rangle \sim 4\%$ and a roughly stable polarization angle of $\langle\Psi_{\rm R\,band}\rangle \sim 15^{\circ}$. Following this ``stable'' phase, the polarization properties of PG\,1553+113 began to vary systematically, with the polarization degree reaching a maximum of $\Pi_{\rm R\,band} \sim 8.7\%$ in June 23 (MJD$\sim$60849). Within approximately two weeks, the polarization degree decreased to $\Pi_{\rm R\,band} \sim 0\%$, accompanied by a large swing in the polarization angle ($\Delta \Psi_{\rm R\,band} > 100^{\circ}$) starting from a polarization angle $\sim45^{\circ}$ down to $\sim-180^{\circ}$ with low amplitude variability. After the R-band polarization degree reached nadir roughly half-way through the rotation, it exhibited a steady, essentially linear increase over the ensuing two months, reaching a pre-rotation level of approximately 8\%. 



 In September 2025, around the time of the third {\it IXPE} pointing  (MJD\,$\sim$\,60930), the optical polarization degree exhibited further strong variations; however, these changes were no longer accompanied by polarization angle swings as pronounced as those observed between June and July (i.e., between MJD\,$\sim$\,60845 and MJD \,$\sim$\,60864).

In the bottom panel of Fig.~\ref{mwl}, we present the contemporaneous {\it Fermi}-LAT light curve \citep[][]{Abdollahi2023}. Around mid December 2025, between MJD 60657 and 60666, PG\,1553+113 displayed enhanced $\gamma$-ray emission, after which the flux decreased significantly. The optical emission seems to have followed this trend (see the top panel in Fig.~\ref{mwl}), such that the first {\it IXPE} observation in 2025 occurred while the source was in a low flux state. 

In the radio, observations at 14~GHz, 17~GHz, and 225~GHz show a comparable polarization degree at $\sim2$--3\% through out the {\it IXPE} observations, although much less-sampled compared to the optical observations. Our campaign also included observations at 36~GHz however, the analysis yielded only upper limits. Close in time to the second IXPE observation, there is a significant drop of the polarization degree to $\sim1.5\%$ at 17~GHz and 0.85\% at 225~GHz. This is in agreement with the optical observations, which show their lowest polarization degree values at around the same time, i.e., while the rotation of the optical polarization angle was taking place. The radio polarization angle remains consistent across all frequencies and roughly perpendicular to the jet axis, except during the second IXPE observation. During the optical rotation the 225\,GHz polarization angle becomes near-parallel to the jet axis.

\section{Discussion and Conclusions}
We have presented the results of a multi-wavelength polarimetric campaign on the blazar PG\,1553+113, centered on three \textit{IXPE} observations performed at distinct flux and polarization states. The source exhibited substantial variability across all wavelengths, characterized by X-ray flux changes by up to a factor of $\sim 5$ and a complex optical polarimetric behavior, including a large EVPA rotation ($\Delta\Psi_{\rm opt} \gtrsim 150^{\circ}$) and a transient drop in the polarization degree to nearly zero. 

\subsection{X-ray Polarization and Jet Geometry}

This behavior highlights the peculiar nature of PG\,1553+113 compared to the broader class of HSP blazars. In the standard scenario for HSPs, the magnetic field is typically compressed by shocks oriented transverse to the jet axis, resulting in a mean field perpendicular to the axis and, consequently, an EVPA aligned with the jet direction \citep{Laing1980, Cawthorne1990, Marscher2008}. This paradigm has been largely confirmed by \textit{IXPE} observations of sources like Mrk\,501 \citep[e.g.,][]{Liodakis2022,Chen2024}.

However, PG\,1553+113 deviates significantly from this trend. As imaged by the MOJAVE program at 15\,GHz, the source is extremely core-dominated, exhibiting only faint extended structure at a projected position angle (PA) between $30^{\circ}$ and $50^{\circ}$ \citep{Lico2014}. Our measured $\Psi_{\rm X} \sim 74^{\circ}\pm9^{\circ}$ in Obs.\,3 is thus oblique to this parsec-scale jet direction, with an offset of $\sim 30^{\circ}$--$45^{\circ}$. This configuration is reminiscent of the `oblique' polarization seen in Mrk\,421 during its first \textit{IXPE} epoch \citep{DiGesu2023}, suggesting that the emission might originate in regions where the magnetic field is either helical or influenced by oblique shocks.

Crucially, the jet geometry in PG\,1553+113 is known to be non-stationary. \citet{Lico2020} reported that the jet position angle varies over time, and \citet{Capecchiacci2025} quoted a jet direction of $\sim 90^{\circ}$, potentially resulting from a precessing jet nozzle or a complex helical motion of the emitting plasma. If we adopt this more recent estimate, our measured $\Psi_{\rm X} = 74^{\circ} \pm 9^{\circ}$ would appear much closer to alignment with the jet axis, shifting the interpretation back toward a shock-compressed field scenario. This would possibly suggest that the previously observed ``obliquity'' of the X-ray polarization may be a transient effect tied to some structural evolution (e.g., precession) of the jet itself, rather than a static property of the source.

\subsection{Possible decoupling between Flux and Polarization}

Our results suggest a complex relationship between X-ray flux and polarization that may challenge the trends typically observed in HSP blazars (see, e.g., Fig.~\ref{mwlzoom}). To date, \textit{IXPE} observations of bright HSPs, such as Mrk\,421 and Mrk\,501, have generally shown that higher X-ray flux states are accompanied by well-defined, significant polarization signals. In PG\,1553+113, however, the degree of magnetic field ordering appears to be independent of the total energetic output of the jet. While one should remain cautious given the $1\sigma$ uncertainties, a comparison between Obs.\,1 and Obs.\,2 is particularly interesting. Despite an increase in the X-ray flux by a factor of approximately two during Obs.\,2, the polarization signal becomes undetectable, failing to reach even a $1\sigma$ threshold. In contrast, the lower-flux state of Obs.\,1 tentatively suggests a polarized signal with $\Pi_{\rm X} \sim 18\%$, being detected up to $\sim2.3\sigma$. This may indicate that the physical conditions responsible for increasing the flux (e.g., a higher efficiency of particle injection or stronger Doppler boosting) do not necessarily enhance, and may even disrupt (e.g., via turbulence), the magnetic field order. 

Furthermore, the non-detection in Obs.\,2 coincides with a period of nearly-zero optical polarization (see Fig.~\ref{mwlzoom}). This synchronized depolarization across all radio, optical, and X-ray bands is a remarkable feature, suggesting a high degree of co-spatiality between the electron populations responsible for the synchrotron emission at the various energy scales. Contrary to other HSPs, where the optical emission often originates from a more extended and turbulent region compared to the compact X-ray site, in PG\,1553+113 the X-ray and lower-energy emitting zones appear to `communicate'  more directly. In this framework, the global depolarization observed during Obs.\,2 may reflect a large-scale stochastic reconfiguration or a surge of turbulence that simultaneously affects the entire synchrotron emitting volume, regardless of the electron energy.

\subsection{The extraordinary optical polarization swing of PG\,1553+113}

The dramatic optical polarization event observed between June and July 2025, characterized by a rapid EVPA rotation exceeding $180^{\circ}$ and a simultaneous drop in $\Pi_{\rm opt}$, represents one of the most extreme polarimetric variations ever sampled for this source. While EVPA rotations are well documented in blazars \citep[e.g.,][]{Abdo2010,Blinov2015, Blinov2016}, the rotation reported here, along with the historical swings discussed by \citet{Raiteri2017}, identifies PG\,1553+113 as a source capable of some of the largest EVPA rotations ever recorded in the literature. These swings in the EVPA can be seen as the signature of a complex magnetic field structure within the jet. In particular, such rotations favor a scenario in which an ordered component, most plausibly a helical magnetic field, is combined with a stochastic component arising from shocks, plasmoid formation, and turbulence.\\
\indent A similar large-scale rotation was observed in 2023 \citep{Middei2023}, although in that instance the degree of X-ray polarization remained remarkably constant throughout the optical EVPA swing, and the swing seemed to also be connected to observed $\gamma$-ray brightening. The authors explained this behavior in light of various scenarios like kink-driven magnetic reconnection \citep{Bodo2021} and the merging of reconnection plasmoids \citep{Hosking2020}. However, the former scenario typically predicts similar polarization degrees across bands, failing to explain the observed chromaticity, while the latter involves timescales much shorter than the $\sim$7 days observed for the swing. Alternatively, the rotation was also interpreted within the Turbulent Extreme Multi-Zone (TEMZ) framework \citep{Marscher2017} as a stochastic random walk, where the lower optical $\Pi$ stems from a more extended and turbulent emission region compared to a more ordered and compact X-ray core.

The 2025 multi-wavelength campaign reveals a spectro-polarimetric behavior that significantly differs from previous epochs. The lack of a significant X-ray polarization detection in Obs.\,2, contrasted with the recovery in Obs.\,3 and the tentative signal in Obs.\,1, may hint at a highly variable emitting environment. In the following, we discuss two possible frameworks aiming at explaining the observed multi-wavelength spectro-polarimetric properties of the 2025 campaign.

\noindent \textbf{Two-Component Model and Vector Cancellation:} 
    We consider two spatially distinct emitting regions (or substructures) within the jet, each characterized by a relatively ordered magnetic field, but with EVPAs differing by somewhat less than $90^{\circ}$. In this framework, the observed rotation and deep depolarization would not be caused by a physical rotation of the field, but rather by the temporal variation in the relative flux contributions of these two components. As one component fades and the other rises, their competing electric vectors increasingly cancel each other, naturally leading to the observed $\Pi_{\rm opt} \sim 0\%$ state. The X-ray polarization behavior would then depend on which component dominates the high-energy synchrotron emission at a given epoch. The nearly orthogonal EVPAs suggested by the data between Obs.\,1 and Obs.\,3 are qualitatively consistent with this ``vector cancellation'' picture, and point to a dynamically evolving magnetic topology within the inner jet.\\
\noindent \textbf{Shock-Turbulence Interaction in a Helical Field:} 
     Alternatively, the observed behavior could be explained by the interaction between a transverse moving shock and a stationary (standing) shock within the jet. The collision between these two structures would trigger a simultaneous increase in the radio, optical, and X-ray fluxes. However, the interaction likely develops strong, localized turbulence, which causes significant depolarization and the observed EVPA rotation during IXPE Obs.\,2, as supported by the optical data (see Fig.\,\ref{mwlzoom}, top middle panel) and the lack of a significant X-ray polarization detection.  This turbulent phase appears to be short-lived, as suggested by the optical polarization degree dropping close to zero. By the time of IXPE Obs.\,3, the moving shock has likely passed the standing shock and becomes more prominent. This leads to the recovery of the characteristic polarization behavior observed in HSP blazars \citep[e.g.,][]{Liodakis2022}, where the polarization degree follows the hierarchy $\Pi_{\rm X}\,>\,\Pi_{\rm opt}\,>\,\Pi_{\rm rad}$ and the X-ray EVPA aligns closely with the jet direction.
In both scenarios, the data support an energy-stratified jet structure in which different electron populations, possibly located in partially overlapping regions, contribute to the optical and X-ray synchrotron emission. These interpretations remain tentative, as the rapid variability and the $1\sigma$ nature of some X-ray constraints do not yet allow us to uniquely identify the underlying engine of the observed rotation.

Finally, we note a key difference when comparing our data with the historical events reported by \citet{Raiteri2017}, who also measured a long optical polarization swing of $\sim350^{\circ}$. Interestingly, their sampled optical EVPA swings were not accompanied by a clear or dramatic depolarization in the same band (see their Fig.\,4), contrary to our 2025 event, which exhibited a near-total drop of $\Pi_{\rm opt}$ to zero during the rotation.

\subsection{Long-term Stability vs. Short-term Variability}

Despite the short-term variations observed in the optical band, the X-ray polarization measured in 2025 remains statistically consistent with the values reported by \citet{Middei2023}. Specifically, both the polarization degree ($\Pi_{\rm X}$) and the polarization angle ($\Psi_{\rm X}$) are compatible between the two epochs. This consistency, also observed in Mrk\,501 \citep[][]{Chen2024}, suggesting that the X-ray emission originates from a persistent acceleration zone, likely associated with the most internal and ordered regions of the jet \citep[e.g.][]{Tavecchio2022}. While the optical emission is susceptible to local, stochastic fluctuations, often attributed to smaller turbulent cells or transient shocks \citep{Marscher2008, Marscher2014}, the X-ray emitting particles probe a ``magnetic backbone'' that may represent a large-scale, ordered magnetic field structure defining the core geometry of the inner jet.\\
\indent Future monitoring with higher cadence and improved sensitivity will be crucial to disentangle the physical origin of the observed polarization variability. In particular, distinguishing between stochastic processes, such as those described within the TEMZ framework, and more deterministic scenarios involving shocks propagating through an ordered magnetic field, will require resolving the detailed temporal evolution of both the polarization degree and angle. These scenarios are expected to produce qualitatively different signatures, such as random-walk behavior versus smooth, coherent rotations.
\\
\subsection{X-ray and optical polarization properties as a function of $\gamma$-ray activity}
\begin{figure*}
    \centering
    \includegraphics[width=0.98\textwidth]{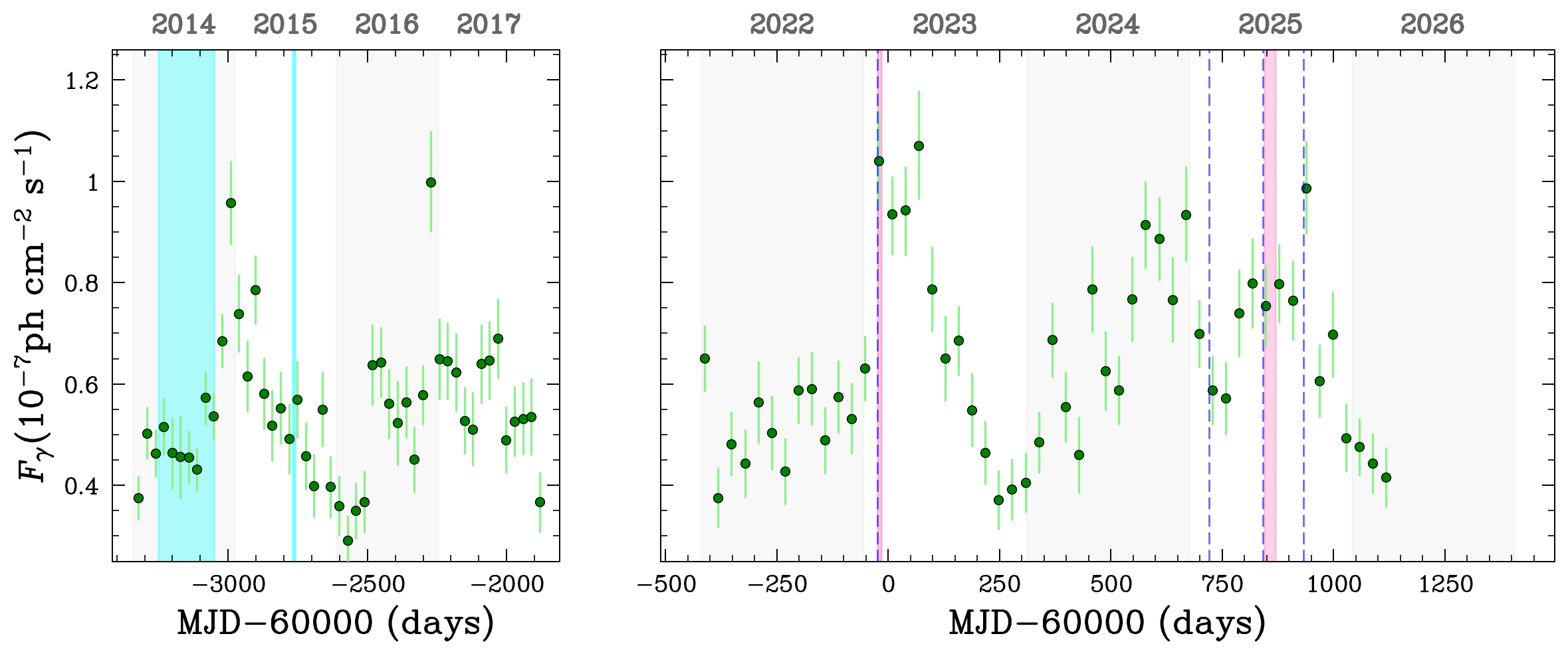}
    \caption{Selected intervals of the monthly \textit{Fermi}-LAT light curve for PG\,1553+113. Vertical blue dashed lines indicate the epochs of the \textit{IXPE} observations, while shaded colored areas in cyan and pink highlight the EVPA swings reported in \citet{Raiteri2017} and \citet{Middei2023}, respectively. Light gray bands are used to mark the alternation of different years.}
    \label{longterm}
\end{figure*}

In Fig.~\ref{longterm} we show two intervals of the {\it Fermi}-LAT light curve of PG\,1553+113, with vertical lines marking the epochs of X-ray observations performed with {\it IXPE} \citep{Middei2023}, including those presented in this work.
Previous studies \citep[e.g.,][]{Ackermann2015} have reported hints of nearly zero-lag correlations among the $\gamma$-ray, X-ray, and optical bands, together with a possible delay of $\sim$\,200 days with respect to the radio emission, suggesting a rather coherent multi-wavelength variability pattern. Although the current dataset does not allow us to establish a clear correlation between the X-ray polarization degree or EVPA swings and the long-term variability pattern, we note that all {\it IXPE} observations with a significant polarization detection (i.e., at $\geq 3\sigma$ confidence level) occurred approximately simultaneously with episodes of enhanced $\gamma$-ray activity (see Fig.~\ref{longterm}). The 2023 observation and the one in September 2025 display a comparable enhanced $\gamma$-ray flux level and strikingly similar X-ray polarization properties. On the other hand, only tentative polarization was observed during {\it IXPE} observation~1 (corresponding to a relatively low $\gamma$-ray flux state), while no clear polarization was detected in X-rays during periods of intermediate-to-high $\gamma$-ray flux ({\it IXPE} observation 2). This temporal behavior may suggest that the conditions leading to detectable X-ray polarization, such as an increased ordering of the magnetic field or enhanced particle acceleration, are more likely to be met during high $\gamma$-ray flux states.

A comparison between the $\gamma$-ray light curve and the epochs of large optical EVPA rotations (Fig.~\ref{longterm}) does not reveal a clear and unique connection between polarization swings and high-energy activity. However, this comparison has a small statistics, being limited  to the EVPA rotations reported in \citet{Raiteri2017}, \citet{Middei2023}, and those identified in this work.

The observed behavior of PG\,1553+113 changes with epoch. During the 2014–2015 flaring event, a large-amplitude EVPA rotation occurred, followed by enhanced $\gamma$-ray activity. Subsequently, in 2015, a smaller variation ($\Delta \Psi_{\rm opt}\,\sim\,60^{\circ}$) was detected after the main flare. Similarly, in 2023 the EVPA rotation appears to precede the increase in $\gamma$-ray flux, suggesting a possible causal link in which the physical conditions leading to the rotation may also trigger enhanced high-energy emission \citep[see, e.g.,][]{Blinov2016}. In contrast, the 2025 rotation occurs during a relatively low $\gamma$-ray state, characterized by a possibly declining flux trend rather than a contemporaneous major flare. This is consistent with recent results on a large number of EVPA rotations in a $\gamma$-ray complete sample of blazars suggesting no correlation between rotation amplitude and $\gamma$-ray brightness (Glykopoulou et al., in prep.). Alternatively, as discussed in \cite{Kiehlmann2017,Blinov2018}, while not all EVPA rotations can be due to random walks of the magnetic field, some of them could. In this case, the TEMZ framework \citep{Marscher2014} would be more consistent with our observations.

These results emphasize the need for sustained polarimetric monitoring of PG\,1553+113 with {\it IXPE}, which represents a unique tool to probe the dynamical evolution of the jet magnetic field and to unveil the physical conditions underlying the connection (or lack thereof) between EVPA rotations and $\gamma$-ray activity, as well as understanding the multi-wavelength quasi-periodic variations of the source brightness.

{\it \textbf{Acknowledgments:}}
We are grateful to the anonymous referee for valuable comments that improved this work. RM warmly thanks Enrico Piconcelli for useful comments and acknowledges financial support from the INAF Scientific Directorate. This work has been partially supported by the ASI-INAF program I/004/11/6. The Imaging X-ray Polarimetry Explorer (\ixpe) is a joint US and Italian mission.  The US contribution is supported by the National Aeronautics and Space Administration (NASA) and led and managed by its Marshall Space Flight Center (MSFC), with industry partner Ball Aerospace (contract NNM15AA18C).  The Italian contribution is supported by the Italian Space Agency (Agenzia Spaziale Italiana, ASI) through contract ASI-OHBI-2022-13-I.0, agreements ASI-INAF-2022-19-HH.0 and ASI-INFN 2024-48-HH.0 ASI-INFN-2017.13-H0, and its Space Science Data Center (SSDC) with agreements ASI-INAF 2025-33-HH.0 and ASI-INFN 2026-5-HH.0, and by the Istituto Nazionale di Astrofisica (INAF) and the Istituto Nazionale di Fisica Nucleare (INFN) in Italy.  This research used data products provided by the IXPE Team (MSFC, SSDC, INAF, and INFN) and distributed with additional software tools by the High-Energy Astrophysics Science Archive Research Center (HEASARC), at NASA Goddard Space Flight Center (GSFC). We acknowledge financial support from ASI-INAF agreement n.\ 2022-14-HH.0.  Some of the data are based on observations collected at the Observatorio de Sierra Nevada; which is owned and operated by the Instituto de Astrof\'isica de Andaluc\'ia (IAA-CSIC); and at the Centro Astron\'{o}mico Hispano en Andaluc\'ia (CAHA); which is operated jointly by Junta de Andaluc\'{i}a and Consejo Superior de Investigaciones Cient\'{i}ficas (IAA-CSIC). The Perkins Telescope Observatory, located in Flagstaff, AZ, USA, is owned and operated by Boston University. The Boston University effort was supported by NASA IXPE Guest Observer grant 80NSSC25K0282. This research was partially supported by the Bulgarian National Science Fund of the Ministry of Education and Science under grants KP-06-H68/4 (2022) and KP-06-H88/4 (2024). The data in this study include observations made with the Nordic Optical Telescope, owned in collaboration by the University of Turku and Aarhus University, and operated jointly by Aarhus University, the University of Turku and the University of Oslo, representing Denmark, Finland and Norway, the University of Iceland and Stockholm University at the Observatorio del Roque de los Muchachos, La Palma, Spain, of the Instituto de Astrofisica de Canarias. The data presented here were obtained in part with ALFOSC, which is provided by the Instituto de Astrof\'{\i}sica de Andaluc\'{\i}a (IAA) under a joint agreement with the University of Copenhagen and NOT. The Submillimeter Array (SMA) is a joint project between the Smithsonian Astrophysical Observatory and the Academia Sinica Institute of Astronomy and Astrophysics and is funded by the Smithsonian Institution and the Academia Sinica. Maunakea, the location of the SMA, is a culturally important site for the indigenous Hawaiian people; we are privileged to study the cosmos from its summit. Partly based on observations with the 100-m telescope of the MPIfR (Max-Planck-Institut f\"ur Radioastronomie) at Effelsberg. Observations with the 100-m radio telescope at Effelsberg have received funding from the European Union's Horizon 2020 research and innovation programme under grant agreement No 101004719 (ORP).  F.E., S.H., J.H., M.K., and F.R. acknowledge support from the Deutsche Forschungsgemeinschaft (DFG, grants 447572188, 434448349, 465409577). This research was funded by the Deutsche Forschungsgemeinschaft (DFG, German Research Foundation) as part of the DFG Research Unit FOR5195 – project number 443220636. G. F. P. acknowledges support by the European Research Council advanced grant “M2FINDERS – Mapping Magnetic Fields with INterferometry Down to Event hoRizon Scales” (Grant No. 101018682). The IAA-CSIC co-authors acknowledge financial support from the Spanish "Ministerio de Ciencia e Innovaci\'{o}n" (MCIN/AEI/ 10.13039/501100011033) through the Center of Excellence Severo Ochoa award for the Instituto de Astrof\'{i}isica de Andaluc\'{i}a-CSIC (CEX2021-001131-S), and through grants PID2019-107847RB-C44 and PID2022-139117NB-C44.  I.L., S.C, A.G., B. A-G., A.F., D.A.L., J.A.K., D.R., O.Z., were funded by the European Union ERC-2022-STG - BOOTES - 101076343. Views and opinions expressed are however those of the author(s) only and do not necessarily reflect those of the European Union or the European Research Council Executive Agency. Neither the European Union nor the granting authority can be held responsible for them.  E. L. was supported by Academy of Finland projects 317636 and 320045. We acknowledge funding to support our NOT observations from the Finnish Centre for Astronomy with ESO (FINCA), University of Turku, Finland (Academy of Finland grant nr 306531). This research has made use of data from the RoboPol program, a collaboration between Caltech, the University of Crete, IA-FORTH, IUCAA, the MPIfR, and the Nicolaus Copernicus University, which was conducted at Skinakas Observatory in Crete, Greece.This work was supported by JST, the establishment of university fellowships towards the creation of science technology innovation, Grant Number JPMJFS2129. This work was supported by Japan Society for the Promotion of Science (JSPS) KAKENHI Grant Numbers JP21H01137. This work was also partially supported by Optical and Near-Infrared Astronomy Inter-University Cooperation Program from the Ministry of Education, Culture, Sports, Science and Technology (MEXT) of Japan. We are grateful to the observation and operating members of Kanata Telescope.

\bibliography{sample701}{}
\bibliographystyle{aasjournalv7}
\appendix
\renewcommand{\thetable}{A}
\section{Swift XRT observations and spectral results}
\begin{longtable*}{l l c c c c c}
\caption{Spectral parameters and fluxes inferred for all the {\it Swift}-XRT observations analysed in the present work. Fluxes are reported in units of
$10^{-11}$\,erg\,cm$^{-2}$\,s$^{-1}$.}\\

\hline
ObsID & Date & $\alpha$ & $\beta$ &
$F_{\rm ixpe}$ & $F_{\rm soft}$ & $F_{\rm hard}$ \\
\hline
\endhead
00015387019 & 2025-01-11 & $2.38 \pm 0.09$ & $0.10 \pm 0.16$ & $1.08 \pm 0.11$ & $1.74 \pm 0.06$ & $1.19 \pm 0.15$ \\
00015387020 & 2025-01-26 & $2.37 \pm 0.20$ & $-0.48 \pm 0.41$ & $1.09 \pm 0.33$ & $1.07 \pm 0.08$ & $1.32 \pm 0.48$ \\
00015387021 & 2025-01-29 & $2.39 \pm 0.14$ & $0.25 \pm 0.26$ & $0.58 \pm 0.10$ & $1.06 \pm 0.05$ & $0.63 \pm 0.13$ \\
00015387022 & 2025-02-04 & $2.32 \pm 0.13$ & $0.25 \pm 0.23$ & $0.47 \pm 0.07$ & $0.78 \pm 0.03$ & $0.52 \pm 0.09$ \\
00015387023 & 2025-02-13 & $2.63 \pm 0.21$ & $-0.39 \pm 0.42$ & $0.55 \pm 0.21$ & $0.86 \pm 0.07$ & $0.63 \pm 0.29$ \\
00015387024 & 2025-02-19 & $2.32 \pm 0.24$ & $0.23 \pm 0.40$ & $0.61 \pm 0.14$ & $1.00 \pm 0.09$ & $0.67 \pm 0.19$ \\
00015387025 & 2025-02-22 & $2.31 \pm 0.12$ & $0.12 \pm 0.19$ & $0.55 \pm 0.07$ & $0.82 \pm 0.03$ & $0.61 \pm 0.09$ \\
00015387026 & 2025-03-03 & $2.25 \pm 0.13$ & $0.68 \pm 0.22$ & $0.51 \pm 0.07$ & $1.05 \pm 0.05$ & $0.54 \pm 0.08$ \\
00015387027 & 2025-03-12 & $2.27 \pm 0.12$ & $-0.10 \pm 0.17$ & $0.84 \pm 0.09$ & $0.99 \pm 0.03$ & $0.96 \pm 0.14$ \\
00015387028 & 2025-03-24 & $2.38 \pm 0.11$ & $0.25 \pm 0.20$ & $0.65 \pm 0.08$ & $1.18 \pm 0.05$ & $0.70 \pm 0.10$ \\
00015387030 & 2025-04-02 & $2.77 \pm 0.80$ & $-0.31 \pm 0.63$ & $0.62 \pm 0.14$ & $1.02 \pm 0.06$ & $0.69 \pm 0.15$ \\
00098116001 & 2025-04-03 & $2.72 \pm 0.14$ & $-0.79 \pm 0.22$ & $0.99 \pm 0.18$ & $1.26 \pm 0.08$ & $1.22 \pm 0.27$ \\
00015387031 & 2025-04-08 & $2.27 \pm 0.17$ & $0.07 \pm 0.30$ & $1.19 \pm 0.25$ & $1.62 \pm 0.10$ & $1.34 \pm 0.36$ \\
00098116002 & 2025-04-10 & $2.61 \pm 0.16$ & $-0.50 \pm 0.27$ & $1.68 \pm 0.32$ & $2.32 \pm 0.16$ & $1.99 \pm 0.48$ \\
00015387032 & 2025-04-11 & $2.20 \pm 0.16$ & $0.11 \pm 0.28$ & $1.23 \pm 0.23$ & $1.58 \pm 0.09$ & $1.37 \pm 0.29$ \\
00015387033 & 2025-04-17 & $2.40 \pm 0.13$ & $-0.32 \pm 0.22$ & $1.28 \pm 0.23$ & $1.54 \pm 0.07$ & $1.50 \pm 0.30$ \\
00098116003 & 2025-04-17 & $2.52 \pm 0.11$ & $-0.12 \pm 0.19$ & $1.33 \pm 0.18$ & $2.19 \pm 0.09$ & $1.49 \pm 0.23$ \\
00015387034 & 2025-04-22 & $2.35 \pm 0.10$ & $-0.15 \pm 0.15$ & $1.93 \pm 0.20$ & $2.47 \pm 0.09$ & $2.21 \pm 0.25$ \\
00098116004 & 2025-04-24 & $2.57 \pm 0.14$ & $-0.36 \pm 0.23$ & $1.45 \pm 0.25$ & $2.12 \pm 0.12$ & $1.68 \pm 0.36$ \\
00015387035 & 2025-04-26 & $2.20 \pm 0.10$ & $0.31 \pm 0.15$ & $1.45 \pm 0.14$ & $2.16 \pm 0.08$ & $1.59 \pm 0.16$ \\
00015387036 & 2025-05-06 & $2.18 \pm 0.19$ & $-0.31 \pm 0.39$ & $2.53 \pm 0.83$ & $2.23 \pm 0.17$ & $3.04 \pm 1.24$ \\
00098116006 & 2025-05-08 & $2.41 \pm 0.07$ & $-0.10 \pm 0.11$ & $2.50 \pm 0.18$ & $3.58 \pm 0.09$ & $2.82 \pm 0.23$ \\
00015387037 & 2025-05-10 & $2.19 \pm 0.09$ & $-0.16 \pm 0.14$ & $2.26 \pm 0.20$ & $2.30 \pm 0.08$ & $2.64 \pm 0.29$ \\
00015387038 & 2025-05-14 & $2.19 \pm 0.09$ & $-0.05 \pm 0.12$ & $1.70 \pm 0.13$ & $1.89 \pm 0.06$ & $1.95 \pm 0.17$ \\
00098116007 & 2025-05-15 & $2.51 \pm 0.13$ & $-0.09 \pm 0.22$ & $0.93 \pm 0.14$ & $1.55 \pm 0.08$ & $1.04 \pm 0.18$ \\
00098116008 & 2025-05-22 & $2.44 \pm 0.11$ & $-0.27 \pm 0.18$ & $1.50 \pm 0.19$ & $1.99 \pm 0.09$ & $1.73 \pm 0.28$ \\
00098116009 & 2025-05-29 & $2.45 \pm 0.13$ & $-0.34 \pm 0.20$ & $1.12 \pm 0.15$ & $1.41 \pm 0.07$ & $1.30 \pm 0.21$ \\
00015387041 & 2025-06-04 & $2.32 \pm 0.10$ & $-0.12 \pm 0.16$ & $1.26 \pm 0.15$ & $1.59 \pm 0.07$ & $1.45 \pm 0.20$ \\
00098116010 & 2025-06-05 & $2.53 \pm 0.10$ & $0.01 \pm 0.17$ & $1.16 \pm 0.13$ & $2.15 \pm 0.08$ & $1.28 \pm 0.15$ \\
00098116011 & 2025-06-12 & $2.58 \pm 0.16$ & $-0.62 \pm 0.27$ & $1.40 \pm 0.28$ & $1.68 \pm 0.10$ & $1.69 \pm 0.39$ \\
00098116012 & 2025-06-19 & $2.73 \pm 0.16$ & $-0.41 \pm 0.31$ & $0.57 \pm 0.11$ & $1.02 \pm 0.06$ & $0.65 \pm 0.14$ \\
00098116013 & 2025-06-26 & $2.78 \pm 0.17$ & $-0.54 \pm 0.37$ & $0.85 \pm 0.48$ & $1.46 \pm 0.09$ & $0.99 \pm 0.40$ \\
00019035001 & 2025-07-05 & $2.33 \pm 0.08$ & $0.02 \pm 0.13$ & $1.37 \pm 0.12$ & $1.94 \pm 0.06$ & $1.54 \pm 0.17$ \\
00019035005 & 2025-07-06 & $2.21 \pm 0.09$ & $0.42 \pm 0.14$ & $1.46 \pm 0.13$ & $2.38 \pm 0.07$ & $1.58 \pm 0.16$ \\
00019035006 & 2025-07-07 & $2.29 \pm 0.09$ & $0.25 \pm 0.15$ & $1.30 \pm 0.12$ & $2.09 \pm 0.07$ & $1.43 \pm 0.14$ \\
00019035007 & 2025-07-08 & $2.35 \pm 0.11$ & $0.06 \pm 0.19$ & $1.32 \pm 0.18$ & $2.00 \pm 0.08$ & $1.47 \pm 0.22$ \\
00019035008 & 2025-07-09 & $2.38 \pm 0.10$ & $0.11 \pm 0.16$ & $0.91 \pm 0.10$ & $1.49 \pm 0.05$ & $1.01 \pm 0.12$ \\
00019035009 & 2025-07-10 & $2.40 \pm 0.12$ & $-0.01 \pm 0.21$ & $0.67 \pm 0.10$ & $1.02 \pm 0.05$ & $0.75 \pm 0.13$ \\
00098116014 & 2025-07-10 & $2.69 \pm 0.15$ & $-0.29 \pm 0.26$ & $0.60 \pm 0.11$ & $1.10 \pm 0.06$ & $0.68 \pm 0.15$ \\
00098116015 & 2025-07-24 & $2.74 \pm 0.13$ & $-0.67 \pm 0.23$ & $0.99 \pm 0.18$ & $1.45 \pm 0.08$ & $1.19 \pm 0.27$ \\
00098116016 & 2025-08-07 & $2.75 \pm 0.12$ & $-0.52 \pm 0.22$ & $0.82 \pm 0.14$ & $1.38 \pm 0.07$ & $0.95 \pm 0.17$ \\
00098116017 & 2025-08-21 & $2.68 \pm 0.15$ & $-0.52 \pm 0.28$ & $0.93 \pm 0.21$ & $1.42 \pm 0.09$ & $1.10 \pm 0.32$ \\
00098116018 & 2025-09-04 & $2.43 \pm 0.11$ & $0.05 \pm 0.18$ & $1.63 \pm 0.20$ & $2.69 \pm 0.11$ & $1.81 \pm 0.26$ \\
00019035010 & 2025-09-11 & $2.26 \pm 0.09$ & $0.19 \pm 0.14$ & $1.78 \pm 0.16$ & $2.62 \pm 0.09$ & $1.97 \pm 0.21$ \\
00019035011 & 2025-09-12 & $2.26 \pm 0.09$ & $0.25 \pm 0.16$ & $1.57 \pm 0.16$ & $2.42 \pm 0.08$ & $1.72 \pm 0.21$ \\
00019035012 & 2025-09-13 & $2.30 \pm 0.08$ & $0.10 \pm 0.14$ & $1.47 \pm 0.14$ & $2.13 \pm 0.07$ & $1.64 \pm 0.17$ \\
00019035013 & 2025-09-14 & $2.25 \pm 0.09$ & $0.13 \pm 0.14$ & $1.47 \pm 0.14$ & $2.05 \pm 0.07$ & $1.63 \pm 0.18$ \\
00019035014 & 2025-09-15 & $2.25 \pm 0.10$ & $0.22 \pm 0.16$ & $1.79 \pm 0.20$ & $2.67 \pm 0.10$ & $1.97 \pm 0.25$ \\
\hline
\label{Swift}
\end{longtable*}

\end{document}